\documentclass[10pt, sigconf,letterpaper,nonacm]{acmart}
\usepackage{caption}
\usepackage{subcaption}
\usepackage{siunitx}
\usepackage{enumitem}
\usepackage{wrapfig}
\usepackage{hyphenat}
\usepackage[utf8]{inputenc}
\usepackage{tikz}
\usepackage{hyperref}

\AtBeginDocument{%
  \providecommand\BibTeX{{%
    \normalfont B\kern-0.5em{\scshape i\kern-0.25em b}\kern-0.8em\TeX}}}

\acmConference{CoNEXT '20}{December 1-4, 2020}{Barcelona, Spain}
\acmPrice{TBD}
\acmDOI{TBD}

\begin{document}

\title{SeQUeNCe: A Customizable Discrete-Event Simulator of Quantum Networks}

\author{Xiaoliang Wu}
\affiliation{%
  \institution{Illinois Institute of Technology}
  \streetaddress{}
  \city{Chicago, IL}
  \country{United States}}
\email{xwu64@hawk.iit.edu}

\author{Alexander Kolar}
\affiliation{%
  \institution{Northwestern University}
  \streetaddress{}
  \city{Evanston, IL}
  \country{United States}}
\email{alexanderkolar2021@u.northwestern.edu}

\author{Joaquin Chung}
\affiliation{%
  \institution{Argonne National Laboratory}
  \streetaddress{}
  \city{Lemont, IL}
  \country{United States}}
\email{chungmiranda@anl.gov}

\author{Dong Jin}
\affiliation{%
  \institution{Illinois Institute of Technology}
  \streetaddress{}
  \city{Chicago, IL}
  \country{United States}}
\email{dong.jin@iit.edu}

\author{Tian Zhong}
\affiliation{%
  \institution{University of Chicago}
  \streetaddress{}
  \city{Chicago, IL}
  \country{United States}}
\email{tzh@uchicago.edu}

\author{Rajkumar Kettimuthu}
\affiliation{%
  \institution{Argonne National Laboratory}
  \streetaddress{}
  \city{Lemont, IL}
  \country{United States}}
\email{kettimut@anl.gov}

\author{Martin Suchara}
\affiliation{%
  \institution{Argonne National Laboratory}
  \streetaddress{}
  \city{Lemont, IL}
  \country{United States}}
\email{msuchara@anl.gov}

\renewcommand{\shortauthors}{X. Wu, A. Kolar, J. Chung, D. Jin, T. Zhong, R. Kettimuthu, and M. Suchara}

\newcommand{\todo}[1]{{\color{blue}{\textbf{TODO}: #1}}}

\begin{abstract}
  Recent advances in quantum information science enabled the development of quantum communication network prototypes and created an opportunity to study full-stack quantum network architectures. This work develops SeQUeNCe, a comprehensive, customizable quantum network simulator. Our simulator consists of five modules: Hardware models, Entanglement Management protocols, Resource Management, Network Management, and Application. This framework is suitable for simulation of quantum network prototypes that capture the breadth of current and future hardware technologies and protocols. We implement a comprehensive suite of network protocols and demonstrate the use of SeQUeNCe by simulating a photonic quantum network with nine routers equipped with quantum memories. The simulation capabilities are illustrated in three use cases. We show the dependence of quantum network throughput on several key hardware parameters and study the impact of classical control message latency. We also investigate quantum memory usage efficiency in routers and demonstrate that redistributing memory according to anticipated load increases network capacity by 69.1\% and throughput by 6.8\%. We design SeQUeNCe to enable comparisons of alternative quantum network technologies, experiment planning, and validation and to aid with new protocol design. We are releasing SeQUeNCe as an open source tool and aim to generate community interest in extending it.
  
\end{abstract}

\maketitle

\section{introduction}
\label{sec:introduction}

Quantum networks promise to deliver new, revolutionary applications that include distributing cryptographic keys with provable security~\cite{bb84, ekert91}, solving distributed computational tasks with exponential reduction in communication complexity~\cite{Brassard_survey_complexity}, or synchronizing clocks with unprecedented accuracy~\cite{clock_synchronization} to name just a few. Recent breakthroughs in quantum engineering have allowed experimental realizations of quantum network prototypes~\cite{Hensen2015, Valivarthi2016} that are supplemented by commercial efforts in the network security arena~\cite{Stucki_2011}.

Prototypes of metropolitan quantum networks with multiple nodes are currently under construction e.g. in Chicago \cite{ArgonneLoop}, the Netherlands~\cite{Castelvecchi2018}, the United Kingdom~\cite{ukqnet}, and South Korea~\cite{KoreaNetwork}. The most significant remaining engineering challenge is building networks that scale both in the number of users and communication distance. Achieving this goal requires a combination of advances in hardware engineering, standardization of new network architectures, development of robust control plane protocols, and techniques that allow reproducible performance testing.

Quantum network simulations can help in understanding the tradeoffs of alternative quantum network architectures, optimizing quantum hardware, and developing a robust control plane. As the size of experimental networks grows and new protocols and technologies are developed, the need to use simulations to model the behavior and interactions of these complex systems increases. The classical networking community has been relying on network simulators to achieve similar goals, with simulators such as ns-3~\cite{NS-3Simulator} receiving widespread use in academia and industry alike.

This work builds a Simulator of QUantum Network Communication (SeQUeNCe), a customizable discrete-event quantum network simulator that models quantum hardware and network protocols. We introduce a modularized design of the simulator that separates functionality at different network layers into modules, a concept similar to the OSI model in classical networking~\cite{OSI_model}. This modularized design allows the testing of alternative quantum network protocols and hardware models and the study of their interactions. Our simulator design also allows easy customizability. SeQUeNCe is freely available as open source on GitHub~\cite{sequence-github}, allowing users to test the performance of new algorithms, protocols, and devices by implementing new functionality in Python and running one of our built-in benchmarks.

Simulating quantum networks is challenging for three reasons. First, although recent work has provided important insights about future quantum network architectures~\cite{van2014quantum,wehner2018quantum}, the lack of consensus about architectural principles requires abstracting certain details and considering many alternatives. We address this challenge by using a modularized design that allows intermodule communication and is more flexible than the OSI model. Second, quantum network protocols are typically described as algorithms~\cite{barrett_eg, bbpssw}, and significant effort is needed to map each of these algorithms to the correct network layer and define its behavior and interactions with other protocols. Our work translates a comprehensive suite of quantum network protocols into state machines that capture all possible protocol states and interactions. The third set of challenges comes from the fundamental differences between quantum and classical networks. For example, while classical networks use packets, quantum networks carry information inside photons generated at megahertz frequencies~\cite{time-bin}. We designed SeQUeNCe to track millions of events per second, and the most intensive simulations reported in this paper generate approximately two billion events.

Development of quantum network simulators that capture the complexity of full-stack quantum networks started receiving significant attention in the past two years. In addition to SeQUeNCe, two other concurrently developed simulators, NetSquid~\cite{link-layer-protocol} and QuISP~\cite{matsuo2019quantum}, were introduced as software packages in mid-2020. Because all three simulators have different internal structure and differ in key assumptions and implementation details, we strongly believe that comparing the work of the three teams will lead to an exchange of ideas and better understanding of quantum networks. We compare the three simulators in \S\ref{sec:relatedwork}.

The main contributions of this work are fourfold:
\begin{itemize}
\vspace{-10pt}
  \item Design and implementation of a scalable, customizable, discrete-event quantum network simulator, SeQUeNCe, that models the behavior of quantum networks with picosecond precision
  \item Release of the simulator as an open source tool freely available on GitHub~\cite{sequence-github}.
  \item Description of a modularized quantum network architecture, including detailed descriptions, models, and implementations of key protocols in each module
  \item Three representative use cases that demonstrate the functionality of the simulator by modeling a metropolitan quantum network under construction in Chicago
\end{itemize}

The paper is organized as follows. In \S\ref{sec:background} we introduce basic terminology, explain how quantum networks operate, and highlight some of their most prominent features. In \S\ref{sec:design} we introduce simulation requirements and the design of the SeQUeNCe simulator that consists of five elementary modules and a simulation kernel. \S\ref{sec:implementation} describes models and presents detailed descriptions of quantum network hardware and control protocols. Simulation results obtained with the SeQUeNCe simulator are presented in \S\ref{sec:usecases}, and related work is discussed in \S\ref{sec:relatedwork}.

\section{Background}
\label{sec:background}





In recent years, much work has gone into the development of hardware that enables quantum networks~\cite{Simon2017}, understanding potential network architectures~\cite{Kimble_2008}, and finding new applications~\cite{wehner2018quantum}. However, architectural principles and the associated control protocols remain nascent, and significant advances in both traditional networking disciplines and quantum engineering are needed to realize a full-stack quantum internet.

We begin by introducing the basic concepts in quantum communication. Quantum networks transmit information encoded in \textbf{quantum states,} mathematical constructs that yield a probability distribution for the measurement outcomes on a quantum system. For an isolated two-level atom, a quantum state is a complex vector denoting a superposition of the atom in the two energy levels. Such a quantum state can be used to encode quantum information---a \textbf{qubit}---with the two levels representing a "0" (i.e., $|0\rangle$) or "1" ($|1\rangle$). 

Quantum states are operated on by \textbf{quantum gates}. They act on a qubit or multiple qubits and change their quantum state. Quantum gates perform reversible logical operations on qubits, in contrast to classical gates that can be irreversible. An example of a quantum gate is the controlled NOT (CNOT) gate~\cite{NielsenChuang}, which takes 2 qubits as input and flips the second qubit (the target qubit) if and only if the first qubit (the control qubit) is 
$|1\rangle$.

Applying CNOT gates on independent qubits can create \textbf{entanglement}, a multipartite (2 or more particles) quantum state that cannot be expressed as a product of states of individual particles~\cite{EPR}. In other words, when entangled, each particle's state is not independent of the others. A well-known example is a Bell state~\cite{NielsenChuang}: 1/$\sqrt{2}(|01\rangle + |10\rangle)$. Entanglement is a fundamentally unique property of quantum mechanics. It can exist among particles even though they are physically separated in space~\cite{EPR}.

Entanglement can be used for \textbf{quantum teleportation}, a process to transfer an arbitrary quantum state (and the qubit information it encodes) from a sender to a distant receiver~\cite{Teleportation}. In order to teleport quantum information, a high-fidelity entanglement shared between the two parties must be first established; classical communication is then used to complete the teleportation protocol and transfer the quantum state from the sender to the receiver.

Noise and loss can harm the high-fidelity entanglement required for teleportation~\cite{bbpssw}. Thus one must perform \textbf{entanglement purification (or distillation)}, a process that transforms many copies of entangled states (typically of a lesser degree of entanglement) into fewer copies of maximally entangled states, using local quantum gates and classical communication~\cite{PurificationPure, bbpssw}.

Global efforts have been undertaken to physically realize quantum networks. These efforts typically resort to technological platforms based on either ground-based fiber-optic networks~\cite{Peev_2009,Sasaki:11,Stucki_2011,China_2000km_link,Valivarthi2016,Hensen2015} or satellite links~\cite{Yin1140}. The first method can be naturally realized through the use of existing global telecommunication fiber-optic infrastructure. These fibers, while having minimal attenuation at the telecom band, still suffer from transmission loss of approximately 0.2 dB/km~\cite{attenuation}. To date, point-to-point photonic fiber links for quantum key distribution have been experimentally realized, with many metropolitan networks already deployed or currently under construction~\cite{Sasaki:11, Peev_2009, Valivarthi2016, CambridgeQN}. Transmitting quantum information beyond metropolitan distances, however, would require quantum repeater nodes at intervals of a few tens to a hundred kilometers in order to relay the quantum information. Physical realization of a functional quantum repeater is  a subject of active research, but long-distance fiber-based quantum networks still remain undeveloped.

The second way to realize a long-distance quantum internet relies on ground-satellite links, which can connect distant nodes of 1,000 km separation with a single satellite station. Nevertheless,  satellite-based quantum networks have obvious shortcomings, including weather restrictions, intermittent operations, and low throughput. In this paper, we thus focus on simulation of a fiber-based quantum network. However, our simulator can be easily extended for satellite-based quantum networks as well.

The unique properties of quantum physics result in three fundamental differences between quantum and classical networks. First, the no-cloning theorem prevents copying quantum information without destroying the original~\cite{kozlowski2019towards}. Unlike in a classical network, one therefore cannot use an amplifier to regenerate signals on long-distance links. Combined with the unavoidable loss during transmission, directly transmitting quantum information in a scalable quantum network is almost impossible~\cite{QIlimit}. 

The second difference is reliance on quantum entanglement, which does not exist for classical networks. To provide reliable information transmission, quantum networks use entanglement to teleport quantum states or rely on quantum error correction. Although entanglement can be established among qubits, classical users cannot directly observe such states from qubits. Only heralded signals (conveyed as classical information) can determine the current quantum state. The usage of entanglement then relies on both classical and quantum information.

The third fundamental difference is the time sensitivity of quantum networks. Many operations, such as Bell state measurement (BSM), require synchronous operations over long distances. Furthermore, quantum information that decoheres over time cannot be easily refreshed as classical information. The lifetime of quantum information simulated in this work is usually on the order of milliseconds to seconds~\cite{fid_eff, rancic_2017}.
\section{System Design}
\label{sec:design}

The differences between quantum and classical networks call for a flexible and scalable quantum network simulator that allows accurate performance analysis of network architectures. We explore its design in this section.

\subsection{Quantum Network Simulation Requirements}

We desire the following simulator characteristics:

\textbf{Realism of Quantum States:} The simulator must be able to accurately trace quantum states, such as entanglement, as well as their fidelity. Furthermore, states can be encoded as time bins~\cite{Timebin}, in the polarization of light~\cite{Bouwmeester}, or as states in quantum memories~\cite{Morton}. The quality of entanglement is a key quantum network performance metric, and loss and decoherence that affect it must be modeled. In SeQUeNCe, entanglement states are represented as complex arrays and hardware models record entanglement fidelity. 

\textbf{Realism of Timing:} Simulation events must be precisely executed at their respective timestamps with their exact ordering to avoid causality errors. Quantum networks are time-sensitive systems, and the arrival times of photons that encode quantum information determine their identity. In addition, the lifetime of qubits in memories is limited, requiring that certain operations be performed with as low latency as possible. Our simulator operates with picosecond precision.

\textbf{Flexibility:} In order to support development of future quantum networks, the simulator must be able to simulate alternative network architectures and new protocols and applications and allow reconfigurable topologies and traffic traces. To that end, SeQUeNCe uses a modularized design that separates functionality into modules that contain protocols that can be reprogrammed. Quick testing of a large number of scenarios is possible by changing parameters in JSON files.

\textbf{Scalability:} We must be able to perform large-scale studies of wide area networks with many components, as well as track quantum states at the individual photon level. Compared with classical packet-level network simulations, we must track photons generated with megahertz frequencies, increasing the number of simulated events by several orders of magnitude. Although this paper focuses on sequential discrete event simulation, we started exploring efficient parallel simulation methods~\cite{pads_2020} and designed a stand-alone simulation kernel to allow portability to high-performance computing systems.

\subsection{Modularized Design of SeQUeNCe}
\begin{figure}[htbp]
    \centering
    \includegraphics[width=0.8\linewidth]{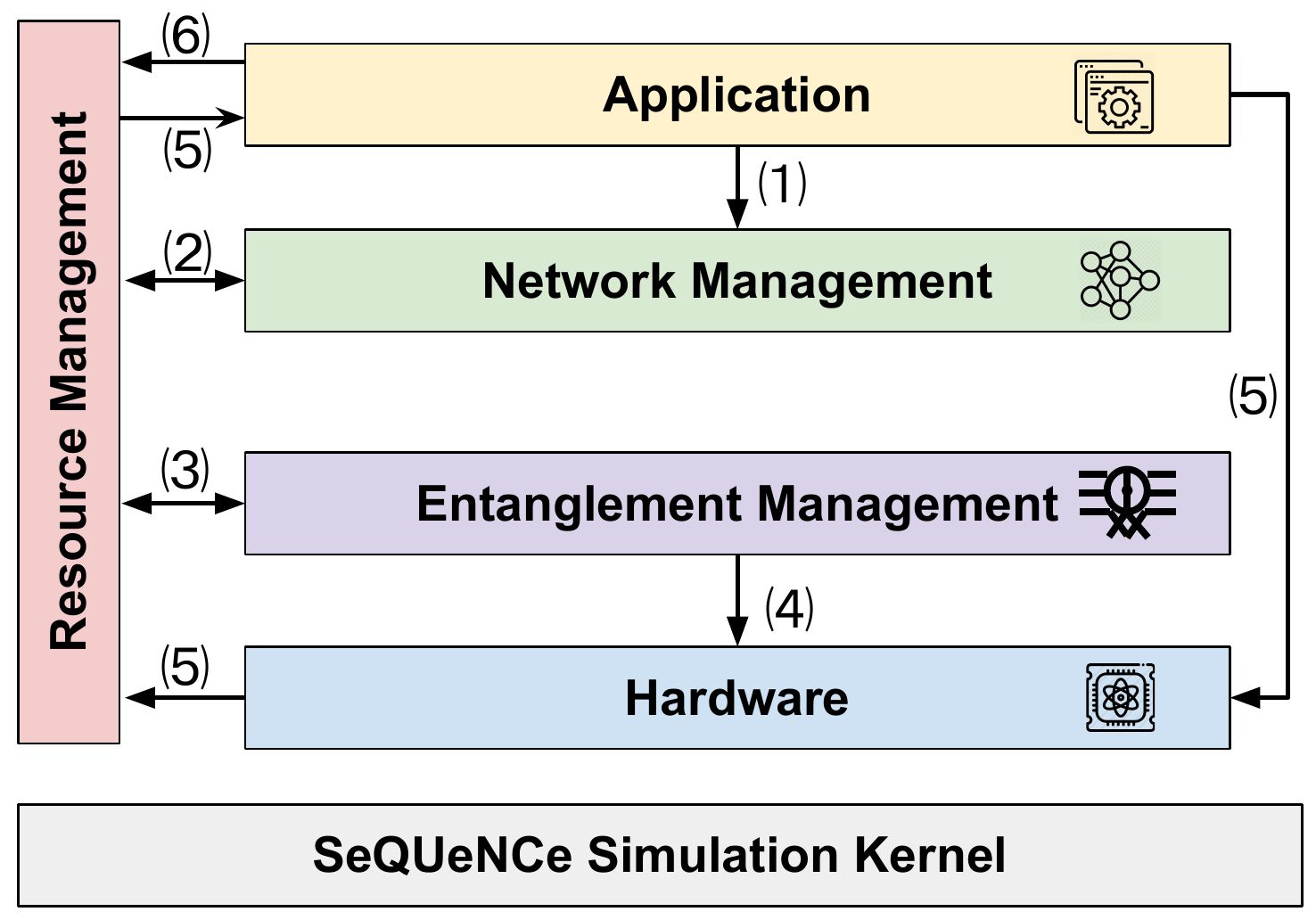}
    \caption{Modularized design of SeQUeNCe closely matching an abstract 
    quantum network architecture.}
    \vspace{-10pt}
    \label{fig:sw_framework}
\end{figure}

To simulate quantum networks, we have to make some assumptions about their architecture. However, quantum network architectures have not been standardized yet, and this topic is the subject of many lively discussions in the recently established Internet Engineering Task Force (IETF) standardization group~\cite{ietf_setup}. We carefully studied the often conflicting quantum network designs, including the seminal works of Rodney Van Meter~\cite{van2014quantum} and Stephanie Wehner~\cite{wehner2018quantum}. To allow simulation of alternative and emerging quantum network architectures, we made minimal assumptions and identified the nascent architectural principles that are common to most quantum network designs. Doing so allowed us to design a simplified quantum network architecture that consists of five modules. SeQUeNCe follows the same modularized design and uses a sixth module, the simulation kernel, to generate events. This design is shown in Figure~\ref{fig:sw_framework}. Next we describe the role and interactions of these six modules.

{}

\textbf{Simulation Kernel} is the heart of SeQUeNCe and enables discrete-event simulation. Simulation time advances in discrete clock ticks, and events generated by the simulation models in all other modules are stored in a priority queue (i.e., min-heap) sorted by the event timestamp. The kernel continuously executes the top event in the heap and advances the simulation time to that particular timestamp. This procedure repeats until the priority queue is empty or a simulation end condition is met. The kernel gives users extreme control over the event execution orders for timing realism and provides interfaces for future parallelized implementation for scalability enhancement.

\textbf{Hardware} module includes models of elementary hardware components used in a quantum network, including quantum channels, classical channels, quantum gates, photon detectors, and quantum memories. Each hardware model provides an interface to allow the Application, Entanglement, and Network modules to query and update its states. Our prior work evaluated the realism of the models of some of these components and demonstrated their interactions~\cite{photonics_19, spw_19}. This work significantly extends these models and introduces new components, such as quantum memories necessary for long-distance quantum communication.

\textbf{Entanglement Management} module includes models of protocols for reliable high-fidelity end-to-end distribution of entangled qubit pairs between network nodes. Specifically, this module includes protocols for entanglement generation, entanglement purification, and entanglement swapping. The role of entanglement generation is to create pairs of entangled qubits. Next, entanglement purification is used to improve the fidelity of the entanglement.  Entanglement swapping is used to transform multiple shorter-distance entangled pairs into a single long-distance entangled pair. SeQUeNCe models these protocols and tracks the quantum state, lifetime, and fidelity of entanglement. Quantum states (including entanglement) are stored as complex arrays of amplitudes together with a list of states that are entangled. This module is allowed to change the quantum state in hardware and release the hardware resource upon completion.

\textbf{Resource Management} module manages local resources within one node. It records the state of the hardware, efficiently allocates resources to applications and entanglement protocols based on commands issued by the network management, and regains control of the hardware with updated states when resources are released. Although the resource manager controls only local resources, the instantiation of entanglement protocols requires cooperation between resource managers on different nodes. This ensures that entangled memories are mapped to the corresponding entangled pairs managed by entanglement protocols.  

\textbf{Network Management} module provides quantum network services based on requests from the local Applications and remote Network Managers. It communicates with the Resource Manager to check the available local resources and generates commands for the Resource Manager to realize an appropriate resource allocation scheme. 

\textbf{Application} module represents quantum network applications and their requests for quantum network resources. Although scientists envision a variety of disparate quantum network applications, including precise clock synchronization, quantum teleportation, or highly secure cryptographic key distribution, all these applications rely on entanglement. An application can initiate distribution of entanglement between another network node and specify the start time, duration, frequency, and fidelity of the entanglement distribution.

The SeQUeNCe simulator is highly reconfigurable. The user is allowed to specify the network topology and a variety of network and protocol parameters in a JSON file. The JSON file automatically creates and configures the appropriate simulation models. In addition, the modularized design was created for easy extendability and allows advanced users to create their own models of new quantum hardware and network protocols.

Here we show an example sequence of steps required to create a flow for entanglement distribution between node-1 and node-3 in a linear network topology node-1 \textrightarrow{} node-2 \textrightarrow{}  node-3. 
After node-1 requests a reservation, the network utilizes quantum memories to distribute entanglement. This example is illustrated step-by-step with the step number corresponding to the labels in Figure~\ref{fig:sw_framework}:

\begin{enumerate}[leftmargin=*]
    \item The Application on node-1 requests a service from its local Network Manager.
    \item The Network Manager invokes a routing protocol to identify a route and announces the request to the network managers on that route, namely, in node-2 and node-3. The Network Managers verify the availability of memories with their local Resource Managers. The request is served upon approval by all nodes. The Network Manager notifies the Application of the approval or rejection of the request.
    \item The Resource Manager allocates memories for use by the Entanglement Manager.
    \item The Entanglement Manager utilizes the allocated memories to execute entanglement generation, purification, and swapping protocols to establish entanglement between node-1 and node-3. 
    \item The Resource Manager continuously updates the state of quantum memories and allocates memories to the Application.
    \item The Application consumes the entanglement between node-1 and node-3. The Application releases the quantum memory after use. At that point, the released memory can be reused to distribute entanglement again.
\end{enumerate}


\section{Design, Implementation, and Simulation of Modules}
\label{sec:implementation}

This section describes the design and implementation of models that follow the modularized architecture described in 
\S\ref{sec:design}. 
We describe models of the elementary hardware building blocks of quantum networks in
\S\ref{subsec:hardware}.
The Entanglement Management module is discussed in \S\ref{subsec:entangle_manage}. We implemented the Barrett-Kok entanglement generation protocol~\cite{barrett_eg} to generate entanglement between adjacent nodes, the BBPSSW~\cite{bbpssw} protocol to purify entanglement, and a swapping protocol~\cite{Entanglementswap} to extend the distance of entanglement.
Our Resource Management module, described in \S\ref{subsec:resource-mgmt},  consists of a memory manager and a rule manager. We present the Network Management module in \S\ref{subsec:network-mgmt}. It uses both a reservation and a routing protocol to create paths.
In \S\ref{subsec:application}
we describe a simplified Application module that mimics quantum network traffic generated by real applications.

\subsection{Hardware} \label{subsec:hardware}
Here we describe how we model the key hardware elements in a quantum network.
Our models of hardware elements simulate the behavior of the physical system and track its state and operational parameters.

\subsubsection{Quantum and Classical Channels} \label{subsec:channels}

Quantum channels in photonic networks use standard telecommunication fiber to transmit quantum information. Our model of a quantum channel has two functions: \texttt{schedule} and \texttt{transmit}. The schedule function is used to determine the earliest available transmission time for the \texttt{transmit} function. The transmit
function sends a photon to the other end of the channel. We model the propagation delay as ${L}/{c^*}$, where $L$ denotes the \textit{length of fiber} and $c^*$ denotes the \textit{speed of light} in the fiber. 
We model the loss rate of the quantum channel as $10^{-\frac{L \cdot \alpha_o }{10}}$, where $\alpha_o$ is the \textit{attenuation} measured in dB/km. To avoid transmitting multiple photons at the same time over the same channel, the modeled quantum channel uses time-division multiplexing (TDM) and assigns a time of transmission to each photon source. Our simulator synchronizes all photon sources sharing a channel in order to ensure proper spacing of photons from different sources.

The classical channel is used to transmit classical information.
Users can define the delay manually. For simplicity, we assume no-loss and perfect reliability for the classical channel in the current version of the simulator. 

\subsubsection{Single Photon Detectors} \label{subsec:detectors}
A single-photon detector\\(SPD) is used to detect individual photon arrivals. An SPD generates an electrical signal upon absorption of a photon and records its arrival time. The \textit{detector efficiency} $\eta$ is the probability that a photon is successfully detected when it hits the detector. The \textit{detector resolution} determines the precision of the timestamps. The \textit{count rate}
determines the constant cooldown time between detection events.
This ``dead time'' is the inverse of the count rate. Another property of the SPD is the \textit{dark count rate}, giving the average number of false positive detections per second caused by outside photons and electrical noise. We model dark count events as a Poisson process. Within the simulation, detectors can be used in a Bell state measurement device. This component receives photons, directs them to an SPD, and ensures the proper entanglement of the photon sources.
We evaluated the accuracy of the channel and detector models in our previous work~\cite{photonics_19, spw_19}.

\subsubsection{Quantum Memory}
\label{subsubsec:memory}
A quantum memory is used to store quantum information in the form of matter (or stationary) qubits.
In this work, we model single-atom memories~\cite{Rempe}, where the qubit is stored as the spin state of a single atom, atomic defect, or ion~\cite{Rempe, Bernien13, Dibos_2018}. We simulate a quantum network composed of multiple single-atom memories connected by fiber-optic channels.

The matter system of a quantum memory consists of two long-lived, low-lying states $|\uparrow \rangle$ and $|\downarrow \rangle$ and one excited state $| e \rangle$. The ``excite'' operation induces the transformation $|\downarrow \rangle \rightarrow | e \rangle$ and $|\uparrow \rangle \rightarrow |\uparrow \rangle$. The transition $|\uparrow \rangle \leftrightarrow | e \rangle$ is not allowed in the physical system. Along with the transformation ${|\downarrow\rangle \rightarrow | e \rangle}$, one photon entangled with the memory may be emitted.

The modeled quantum memory
has two functions: 
\texttt{excite} and \texttt{expire}. The \texttt{excite} function implements the excite operation described above and may cause the memory to emit a photon. The probability of photon emission is decided by the quantum state and the \textit{efficiency of memory} $e$. Given quantum state $\alpha |\downarrow \rangle + \beta |\uparrow \rangle$ stored in a memory, the probability of emitting a photon is given by $e |\alpha|^2 $. The quantum memory then needs time to relax its quantum state back to the ground state before the next excite operation. This time of relaxation determines the \textit{frequency} of the quantum memory. The excite function is typically used for generating entanglement in conjunction with a BSM (described in 
\S\ref{subsec:detectors}). The generated entanglement has a limited lifetime (the \textit{coherence time}) that starts with the excite operation and ends with a scheduled \texttt{expire} operation that resets the quantum state.

When entanglement is established between the states of two memories, both need to maintain additional information such as the identity of the entangled quantum states. In our implementation, the quantum memory also maintains the fidelity of entanglement, ranging from 0 (no entanglement) to 1 (perfect entanglement). 

We consider a repeater architecture based on single-atom memories (as defined above), where the fidelity of entanglement is dependent on atom-cavity cooperativity---an experimental parameter quantifying the coupling strength between an atom memory and a single photon. Based on~\cite{fid_eff}, we have Equation~\ref{eq:fidelity} for entanglement fidelity, where $C$ denotes the atom-cavity cooperativity, $\gamma$ denotes the bare atom's optical decay rate, $\gamma^*$ denotes the optical pure dephasing rate, and $\Delta_\omega$ denotes the difference between the optical transition frequencies of the two atomic memories to be entangled.

\begin{equation}
    \label{eq:fidelity}
    F_{\mathrm{entangle}} = \frac{1}{2}\left(1 + \frac{(C+1)^2\gamma^2}{\left((C+1)\gamma + 2\gamma^*\right)^2 + \Delta_\omega^2}\right)
\end{equation}

\begin{wrapfigure}{r}{0.27\textwidth}
    \centering
    \includegraphics[width=1.0\linewidth]{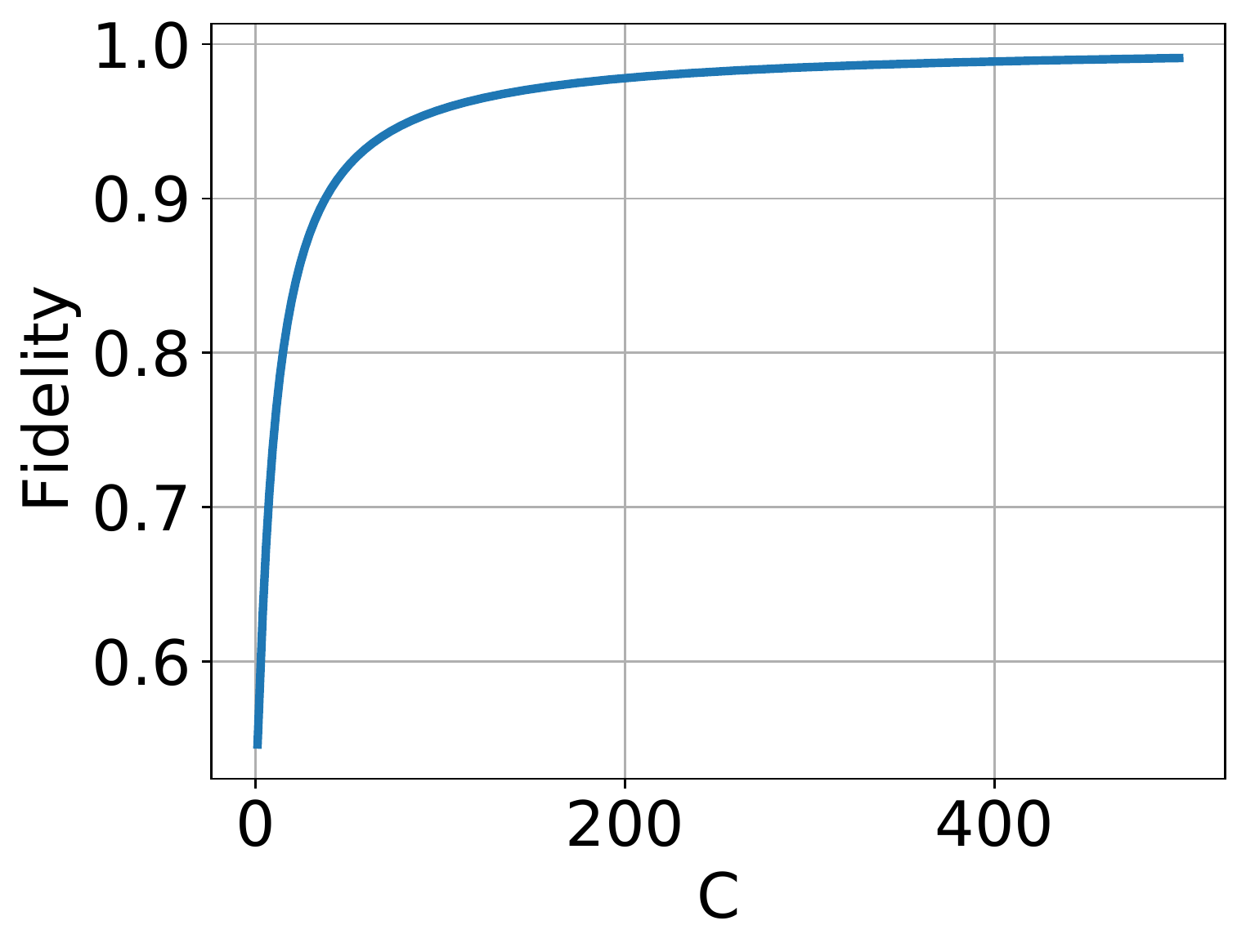}
    \caption{Fidelity of entanglement with varying atom-cavity cooperativity.}
    \vspace{-5pt}
    \label{fig:fidelity_vs_c}
\end{wrapfigure}

Our SeQUeNCe simulator models quantum memories based on single erbium (Er) ions in solids, because Er optical emission is conveniently within the telecommunication C-band and Er has demonstrated spin coherence of over 1 second \cite{rancic_2017}. We choose the following experimental parameters for the simulation: $50 \leq C \leq 500$~\cite{Dibos_2018, raha_2020}, $\gamma = 14$ Hz ~\cite{Bottger_2009, McAuslan_2009}, $\gamma^* = 32$ Hz~\cite{fid_eff}, and $\Delta_\omega = 0$. Figure~\ref{fig:fidelity_vs_c} shows the entanglement fidelity as a function of $C$ according to Equation~\ref{eq:fidelity}.


We also have a relation of the memory efficiency $e$ to the photon collection efficiency $\eta_c$~\cite{fid_eff} as

\begin{equation}
    \label{eq:efficiency}
    e = \eta_c \frac{C}{C+1},
\end{equation}
\noindent where $e \approx \eta_c$ for $C \gg 1$, which is typical in experiments. Realistic values for $e$ range from $10^{-2}$ to $\approx$1 depending on specific photonic coupling techniques used~\cite{Bernien13, Tiecke,Dibos_2018, Zhong_2018}.

\subsubsection{Nodes}
\label{subsec:nodes}


We follow the outline from a recent IETF draft \cite{ietf-principles} and define two types of network nodes. The \textit{router nodes} implement the full-stack functionality described in this section in order to reliably distribute entanglement in the network. 
Our router node has all the functionality of a \textit{quantum repeater} to overcome photon loss described in \S\ref{subsec:channels}. The router also allows routing in the traditional sense. The second type of node is the \textit{BSM node} that is required by the Barrett-Kok entanglement generation protocol described in \S\ref{subsec:entangle_manage}. These nodes are placed in the middle of each link as shown in Figure \ref{fig:bar_init}.

\subsection{Entanglement Management}
\label{subsec:entangle_manage}

The Entanglement Management module includes models of protocols for entanglement generation, purification, and swapping. We reference existing protocols and model their functional behavior, including state machines and the exchange of classical messages. 
The Entanglement Management module provides an interface for instantiating and terminating entanglement protocols, which the Resource Management module utilizes to establish entanglement. It also provides interfaces to receive notifications of events such as photon detection and entanglement expiration, used by the Hardware module. Upon receiving an entanglement expiration notification, the corresponding protocol instances will release their resources and terminate themselves. 


\subsubsection{Barrett-Kok Entanglement Generation Protocol}
\label{subsubsec:barrett}

The Barrett-Kok entanglement generation protocol~\cite{barrett_eg} utilizes matter qubits and linear optics. Compared with the DLCZ entanglement generation protocol~\cite{dlcz}, this protocol can tolerate the most significant hardware errors such as detector loss and spontaneous emission.

\begin{wrapfigure}{r}{0.27\textwidth}
    \centering
    \vspace{-0.5pt}
    \includegraphics[width=1.0\linewidth]{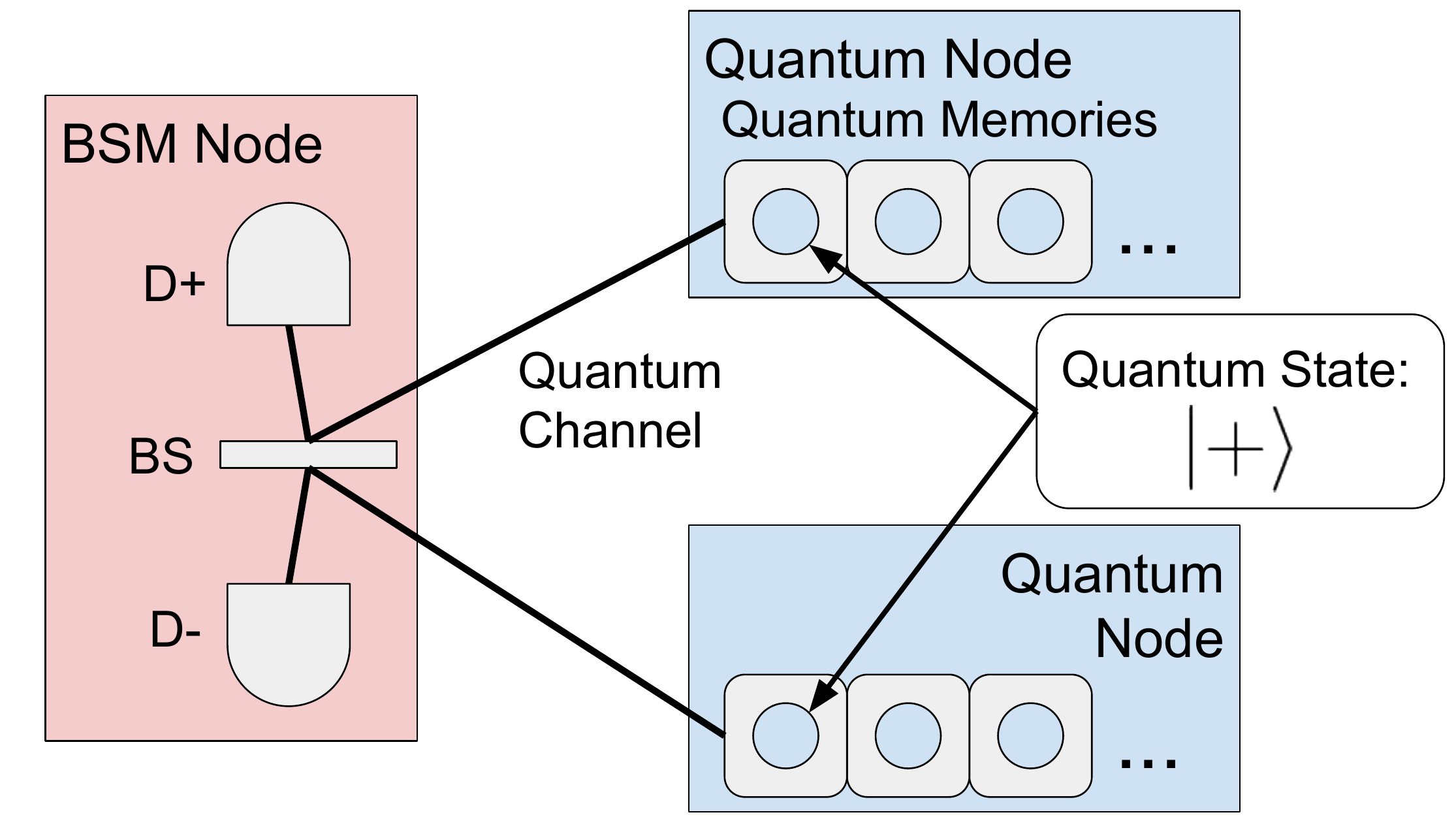}
    \caption{Barrett-Kok generation protocol. Quantum node and BSM node are defined in \S\ref{subsec:nodes}.}
    \vspace{-10pt}
    \label{fig:bar_init}
\end{wrapfigure}

Figure \ref{fig:bar_init} shows the setup of the Barrett-Kok entanglement generation protocol, where two quantum memories are connected to a beam splitter (BS) by optical channels. The quantum memories are prepared in the $| + \rangle = \frac{1}{\sqrt{2}}({|\uparrow \rangle} + |\downarrow \rangle)$ state; the 50:50 BS is used to erase the photon source information; and the single photon detectors $D_+$ and $D_-$ are used to detect emitted photons. Quantum nodes synchronize the times to excite the memories, guaranteeing that the emitted photons from both sides arrive at the BS simultaneously.

The protocol involves two rounds. The first round generates entanglement between two qubits, and the second round verifies the entangled state. The protocol produces a pair of entangled memories with maximal entanglement state $| \Psi^+ \rangle$ or $| \Psi^- \rangle$. For more details on the operation of the protocol as well as our implementation, see Appendix~\ref{apdx:barret}.

\subsubsection{BBPSSW Purification Protocol}

Entanglement purification protocols were developed to address entanglement imperfections caused by factors including imperfections and decoherence in quantum memories. A purification protocol~\cite{entanglement_purification} consumes several pairs of entangled qubits with low quality $F$ to produce a pair of entangled qubits with high quality $F'$. Purification requires the use of quantum gates.

In this work, we implement the BBPSSW protocol~\cite{bbpssw} for purification. The fidelity improvement is shown in Equation~\ref{eq:pur-f} and the success probability in Equation~\ref{eq:pur-psuc}, where $p_{suc}$ denotes the probability of success. For more details on the design and implementation of our BBPSSW protocol model, see Appendix~\ref{apdx:purification}.

\begin{equation}
    \label{eq:pur-f}
    F' = \frac{F^2 + [(1-F)/3]^2}{F^2+2F(1-F)/3+5[(1-F)/3]^2}
\end{equation}

\begin{equation}
    \label{eq:pur-psuc}
    p_{suc} = F^2 + 2F(1-F)/3 + 5([(1-F)/3]^2
\end{equation}

\subsubsection{Swapping Protocol}
\label{subsub:swapping}

\begin{figure}[htbp]
    \centering
    \includegraphics[width=0.8\linewidth]{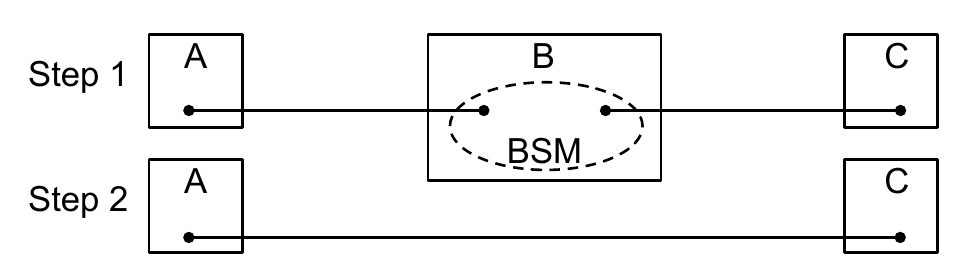}
    \caption{Two short-distance entangled pairs are consumed to distribute long-distance entanglement by the swapping protocol.}
    \vspace{-10pt}
    \label{fig:es_circuit}
\end{figure}

Entanglement swapping is used in quantum networks to extend the distance of entanglement. Figure~\ref{fig:es_circuit} shows the entanglement state before and after the swapping protocol.
To illustrate the procedure, we consider three nodes A, B, and C arranged in a linear topology. Initially, one memory at A and one at C are each entangled to memories at B. Entanglement swapping involves application of a BSM at node B, after which the memories at node A and C are entangled with each other. After this operation, B must send a result message to nodes A and C to determine the exact entanglement state.
The message must include the BSM result, the updated fidelity of entanglement, the updated identity of the entangled memory (which memories are entangled on A and C), and the updated lifetime of entanglement. Note that we choose the smaller value as the updated lifetime if two memories have different lifetimes.

{}

The \textit{success probability} and \textit{degradation} of swapping $F_d$ are defined by the operations performed at the intermediate node.
The degradation comes from imperfect gate operations. Given two pairs of entangled memory with fidelity $F_1$ and $F_2$, the fidelity $F$ of new entanglement after swapping is~\cite{fid_eff}

\begin{equation}
    \label{eq:es_fidelity}
    F = F_1 F_2 F_d .
\end{equation}

For details on the design and implementation of our entanglement swapping protocol model, see Appendix~\ref{apdx:es}.

\subsection{Resource Management} \label{subsec:resource-mgmt}

\begin{figure}[htbp]
    \centering
    \includegraphics[width=\linewidth]{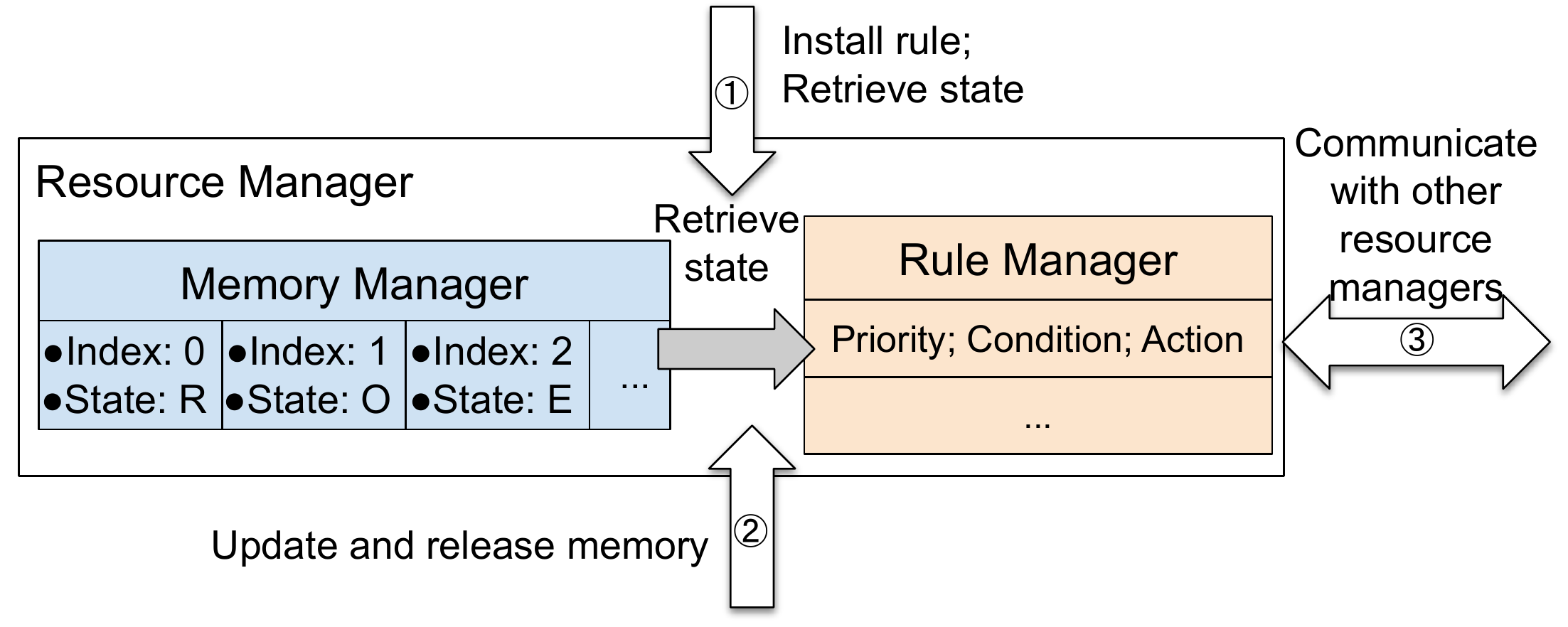}
    \caption{Resource Management module.}
    \label{fig:resource_manager}
    \vspace{-5pt}
\end{figure}

We design and implement a Resource Manager module to manage local resources. Similar to the design of QuISP~\cite{matsuo2019quantum}, the Resource Manager uses an internal set of rules to manage quantum memories. Figure~\ref{fig:resource_manager} shows the structure of the Resource Manager.
The Network Management module and Application module use interface \textcircled{1} to retrieve the state of local resources and/or to install rules. The Hardware and Entanglement Management modules
use interface \textcircled{2} to update the hardware state and acquire/release resources. Interface \textcircled{3} is used to communicate with Resource Managers on other nodes. This communication is used to instantiate and terminate entanglement protocol instances. 
When the Resource Manager creates (or terminates) a protocol instance, it will send a message to the Resource Manager on the remote node(s) to create a (or terminate the) corresponding entanglement protocol instance.

Within the Resource Manager, the Memory Manager traces the states of quantum memories.
The quantum memories can be in one of three states: \texttt{raw}, \texttt{entangled}, or \texttt{occupied}. The \texttt{raw} state means that the memory is not entangled to other memories and is not in use. The \texttt{entangled} state means that the memory is entangled to another memory or memories; entanglement information (such as the identity of entangled memories and fidelity) is also recorded by the memory manager. The \texttt{occupied} state means that the memory has been allocated to an entanglement protocol instance, which  prevents conflict in memory allocation. The transition between states relates to the results of entanglement protocols.

Another component of the Resource Manager is the Rule Manager. The Rule Manager applies a rule set to manage the hardware resources, determining where to allocate them to achieve the desired long-distance entanglement. It installs new rules received from the Network Management module
and uninstalls expired rules. A rule consists of three parts: priority, condition, and action. Rules are sorted by priority from high to low. The condition of a rule determines whether the state of a given memory fits the rule. If the state fits the rule, the action of the rule will allocate the memory to an application or instantiated entanglement protocol. As an example, consider a rule manager containing two rules and one memory. The two rules have the same condition (e.g. the memory must be in the \texttt{raw} state) and different priorities and action (e.g. the action of the rule with high priority allocates the memory to an instance of the entanglement generation protocol; the action of the rule with low priority does nothing). The rule manager first accesses the rule with high priority. Then, the rule manager retrieves the state of the memory from the memory manager. If the state of the memory satisfies the condition of rule (e.g. memory in \texttt{raw} state), the action of the rule is executed (e.g. the rule creates an instance of the generation protocol). If the condition is not met, the rule manager accesses the rule with lower priority to check whether the memory fits the rule.

\subsection{Network Management} \label{subsec:network-mgmt}


The Network Management module enables applications to reserve network resources. This reservation-based approach is inspired by the architecture proposed in a recent IETF quantum internet working group draft~\cite{ietf_setup}. Reservations are necessary to enable better coordination and to conserve the extremely limited quantum memories in emerging quantum network architectures.

To create entanglement between two nodes, the application makes a reservation request consisting of the following 6 elements:
\begin{itemize}[leftmargin=*]
    \item Initiator: the initiator of the entanglement connection that is sending a reservation request (classical message) towards the Responder
    \item Responder: the other end of the connection setup process (and future entanglement connection), where the message sent by the Initiator terminates
    \item Fidelity: the target fidelity of distributed entanglement
    \item Memory size: the memory provided by the Initiator to distribute entanglement and requested of the Responder
    \item Start time: the time from when the resources need to be available for use by the application
    \item End time: the time when the resources can be released 
\end{itemize}

\begin{figure}[htbp]
    \centering
    \includegraphics[width=0.9\linewidth]{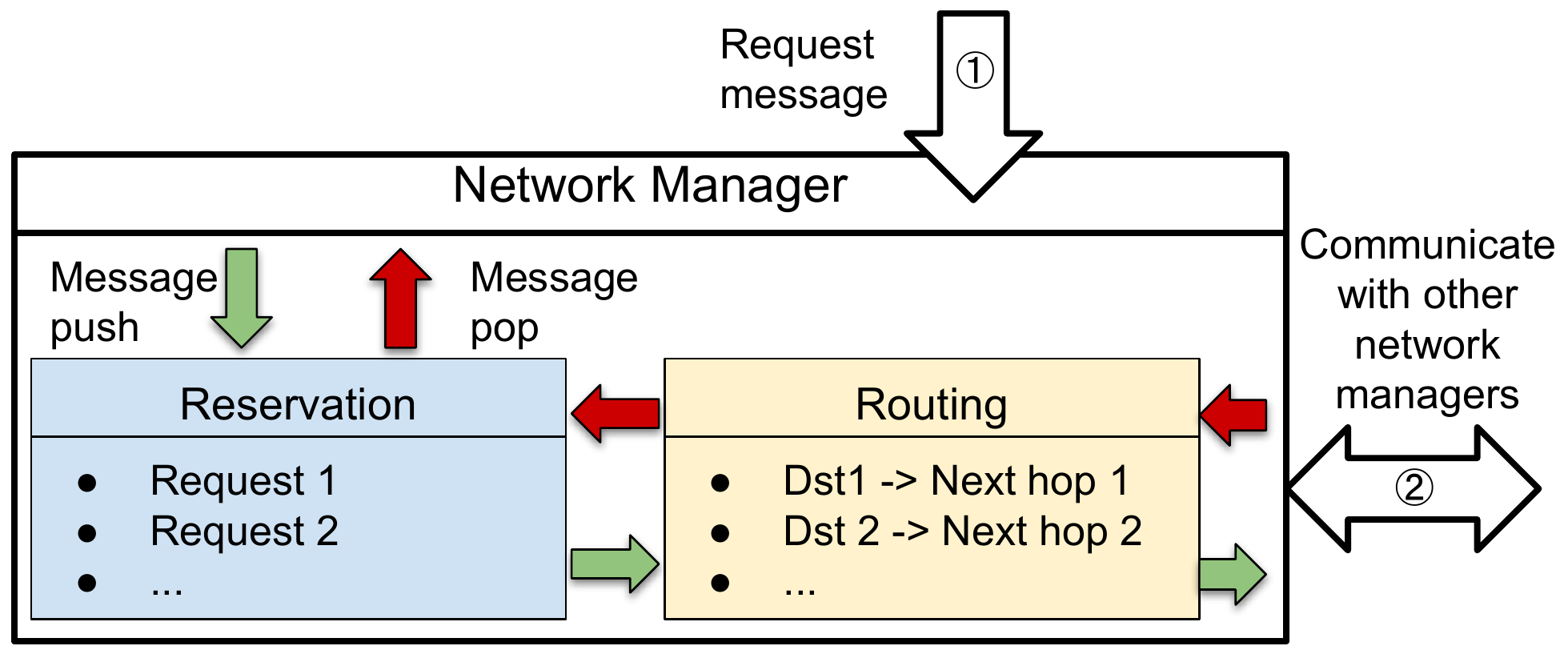}
    \caption{Network Management module.}
    \vspace{-10pt}
    \label{fig:net_manager}
\end{figure}

This module has two duties: reservation and routing. Reservation is responsible for fulfilling reserving the local resources. Routing provides a path of entanglement distribution to satisfy the end-to-end reservation. The reservation and routing protocols are designed to satisfy these duties. Figure~\ref{fig:net_manager} shows the stack structure of the Network Manager. 

Upon receiving a reservation request from an application \textcircled{1}, the Network Manager pushes it to the reservation instance which reserves appropriate resources (if available) and pushes the request to the routing instance.
The routing instance determines the next hop based on its forwarding table (details of this table are given below) and pushes it to the Network Manager, which sends the message to the next hop. We use the length of quantum channels to calculate the shortest path, which provides a static forwarding table for the routing protocol on every node. This method can be updated to use a wide variety of routing protocols including the ones used in classical networks, such as OSPF~\cite{moy1998ospf}.

Upon receiving a message from a neighboring node \textcircled{2}, 
the Network Manager pops it to the routing instance, which then pops it to the reservation. If the reservation instance approves the request, it attaches local information (e.g., the identity of the node) and pushes it back to the routing instance to send the request to the next hop. If the message is rejected or it reaches the Responder, the decision of the Resource Manager will be sent back on the reverse path. If the reservation returns approved, all nodes prepare their local rules to match the reservation. When the result arrives at the Initiator, the reservation instance pops it to the Network Manager, which then pops it to the Application. 
A benefit of this stack structure is that it is easily expanded. For example, an authentication protocol could be placed at the top of the reservation protocol to improve the security of a network while the rest of the protocols are inherited.
For more details about the design of the reservation protocol, see Appendix~\ref{apdx:reservation}.

\subsection{Applications} \label{subsec:application}
The role of the Application module is to consume entanglement. Quantum network applications are heterogeneous (with varying requirements for throughput, fidelity, etc.), so in this work we create an abstraction of a single application with random requests. These requests randomly choose a Responder node (with the Initiator assumed to be the host node) as well as a target fidelity for entanglement. Also assigned randomly are the number of memories and the duration of the reservation. If a request fails, only the Responder and fidelity are kept;  the other parameters are randomly reassigned, and the updated request is sent after waiting for one second. We record the results of simulations using this application description in \S\ref{sec:usecases}.

\section{Three Simulation Use Cases}
\label{sec:usecases}

This section highlights three representative examples of simulations performed with SeQUeNCe. The intent of these use cases is not to focus on extensive quantum network performance evaluations, work that is far beyond the scope of a single research paper or single research group. Instead, these and many other examples are part of the open-source release of SeQUeNCe to help the research community start using the tool. In addition, this paper makes several interesting observations that include quantifying the impact of quantum memory parameters, the adverse effect of \emph{classical} channel delays on \emph{quantum} channel throughput, and the impact of memory distribution policy on memory utilization and overall network performance.

\subsection{Simulation Setup}
\begin{figure}[htbp]
    \centering
    \includegraphics[width=\linewidth]{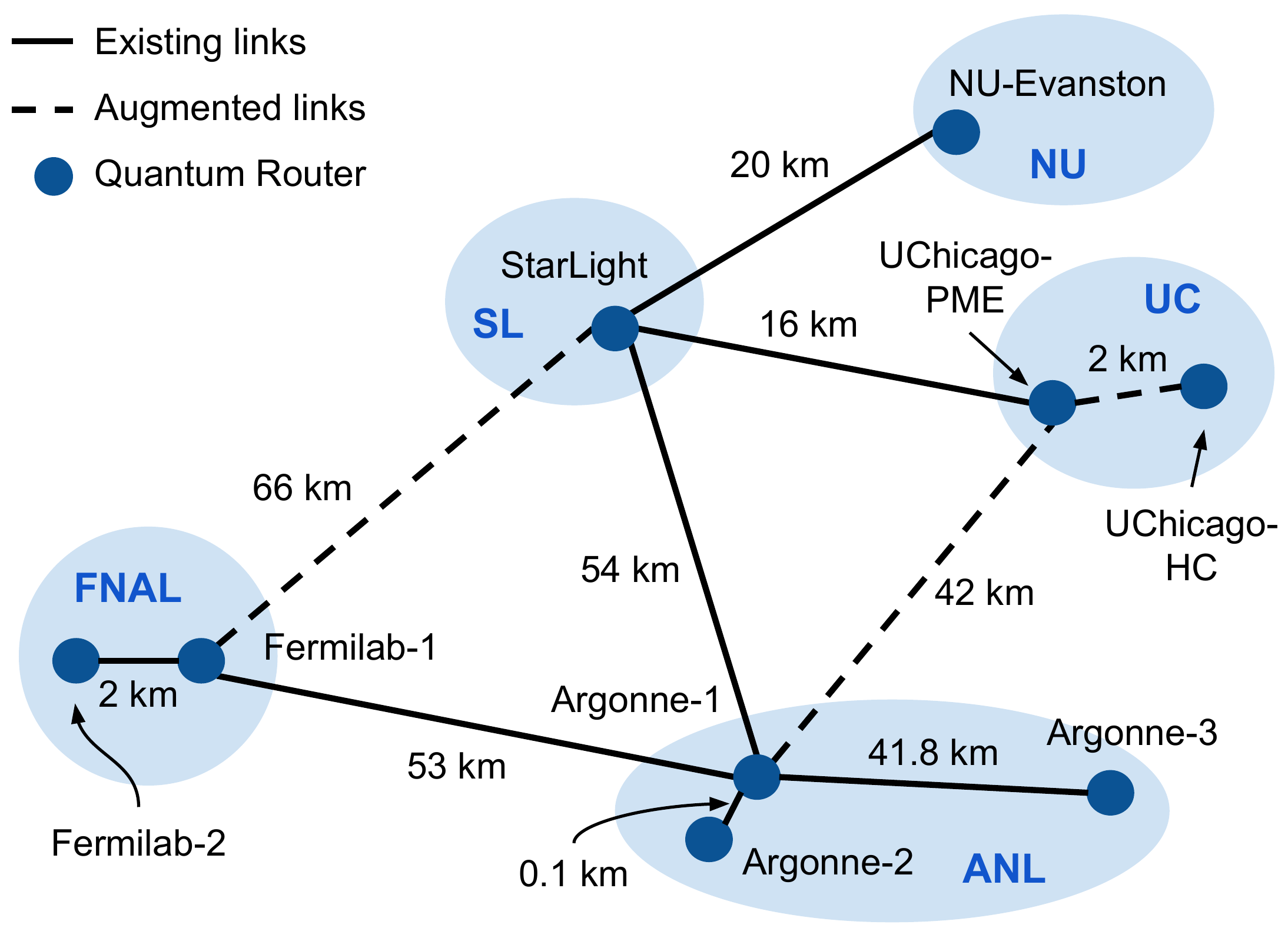}
    \caption{Chicago metropolitan quantum network topology. Solid lines represent existing quantum links. Dashed lines represent augmented quantum links added to improve connectivity.}
    \vspace{-10pt}
    \label{fig:cqx-topo}
\end{figure}

Our use cases share the following simulation setup. The topology is modeled after the Chicago metropolitan quantum network, which consists of 9 nodes located at 5 sites as depicted in Figure~\ref{fig:cqx-topo}. The simulated topology augments the actual topology by adding three quantum links to provide richer connectivity. In addition, a BSM node required by the generation protocol (see \S\ref{subsubsec:barrett}) is placed in the middle of every optical link to enable BSM operations.

{}

The quantum channel delay is modeled as the propagation delay on the optical fiber (see \S\ref{subsec:channels}). The classical channel delay was obtained by performing round-trip time (RTT) measurements between the 5 sites, as shown in Table~\ref{tab:rtt}. The one-way classical channel delay was calculated by halving and averaging the two-directional RTTs between sites, and one-way delays within a site were set to 0.25 ms. Each of the 9 nodes in our network is equipped with quantum memories and supports all protocols described in \S\ref{sec:implementation}. Nodes are therefore able to assume the role of a quantum router. The default hardware parameters in our setup are shown in Table~\ref{tab:parameters}. These values represent properties of state-of-the-art hardware components.

\begin{table}[]
\caption{RTT delays between all pairs of network sites. Rows represent the sender, and columns represent the receiver.}
\vspace{-5pt}
\begin{tabular}{|l|l|l|l|l|l|}
\hline
& \textbf{ANL}     & \textbf{FNAL}     & \textbf{NU}       & \textbf{SL}      & \textbf{UC}      \\ \hline
\textbf{ANL} & --     & 2.65 ms & 1.95 ms  & 1.76 ms & 5.20 ms \\ \hline
\textbf{FNAL} & 2.62 ms & --     & 2.91 ms  & 2.67 ms & 3.84 ms \\ \hline
\textbf{NU}  & 1.99 ms & 2.91 ms & --      & 0.80 ms & 3.83 ms \\ \hline
\textbf{SL}  & 1.77 ms & 2.70 ms & 0.79 ms  & --     & 3.30 ms \\ \hline
\textbf{UC}  & 2.94 ms & 3.94 ms & 6.755 ms & 2.99 ms & --     \\ \hline
\end{tabular}
\label{tab:rtt}
\end{table}

In the course of simulations, applications running on all nodes request quantum network resources repeatedly for 1,000 seconds of simulation time, and performance metrics such as entanglement throughput and memory utilization are recorded. The application requests are random, choosing a random Responder, random target entanglement fidelity between 0.8 and 1, duration of the reservation between 10 and 20 seconds, and a start time at 1 to 2 seconds after the present time. These requests reserve a number of memories between 10 and half of the available memory capacity in the initiator node. As described in \S\ref{subsec:application}, these random requests model the heterogeneous requirements of quantum network applications.

\begin{table}[]
\caption{Parameters of hardware models and their default values.}
\vspace{-5pt}
\begin{tabular}{|l|l|}
\hline
\textbf{Parameter}                                                          & \textbf{Value}      \\ \hline
Memory efficiency \S\ref{subsubsec:memory}                      & $e = 0.75$          \\ \hline
Memory frequency  \S\ref{subsubsec:memory}                             & $f_m = 20$ kHz      \\ \hline
Memory coherence time  \S\ref{subsubsec:memory}                    & $t_c = 1.3$ s       \\ \hline
Atom-cavity cooperativity \S\ref{subsubsec:memory}                           & $C = 500$           \\ \hline
Memory array size                                  & $a = 50$            \\ \hline
Detector efficiency    \S\ref{subsec:detectors}                        & $\eta = 0.8$        \\ \hline
Detector count rate    \S\ref{subsec:detectors}                   & $r = 50$ MHz        \\ \hline
Detector dark count    \S\ref{subsec:detectors}          & $d  \approx 0$ /s          \\ \hline
Detector resolution    \S\ref{subsec:detectors}                   & $s = 100$ ps        \\ \hline
Attenuation      \S\ref{subsec:channels}              & $\alpha_o = 0.2$ dB/km   \\ \hline
Speed of light    \S\ref{subsec:channels}                              & $c^* = 2 \times 10^8$ m/s \\ \hline
Channel TDM time frame   \S\ref{subsec:channels}                      & $t_f = 20$ ns            \\ \hline
Gate fidelity    \S\ref{subsub:swapping}                                            & $F_{gate} = 0.99$                \\ \hline
Swap success probability   \S\ref{subsub:swapping}                  & $p_{swap} = 0.64$         \\ \hline
\end{tabular}
\label{tab:parameters}
\vspace{-10pt}
\end{table}

\subsection{Comparison of Quantum Memory Parameters} \label{subsec:impact-of-qmem}

The attributes of quantum memories greatly affect network performance. As Equation~\ref{eq:fidelity} shows, fidelity of entanglement increases with increasing atom-cavity cooperativity of the quantum memory. Accordingly, we evaluated the network performance for different combinations of cooperativities and efficiencies. We assumed that the memory frequency is 2 KHz and the target fidelity is between 0.8 and 1.

\begin{figure}[]
    \centering
    \includegraphics[width=0.8\linewidth]{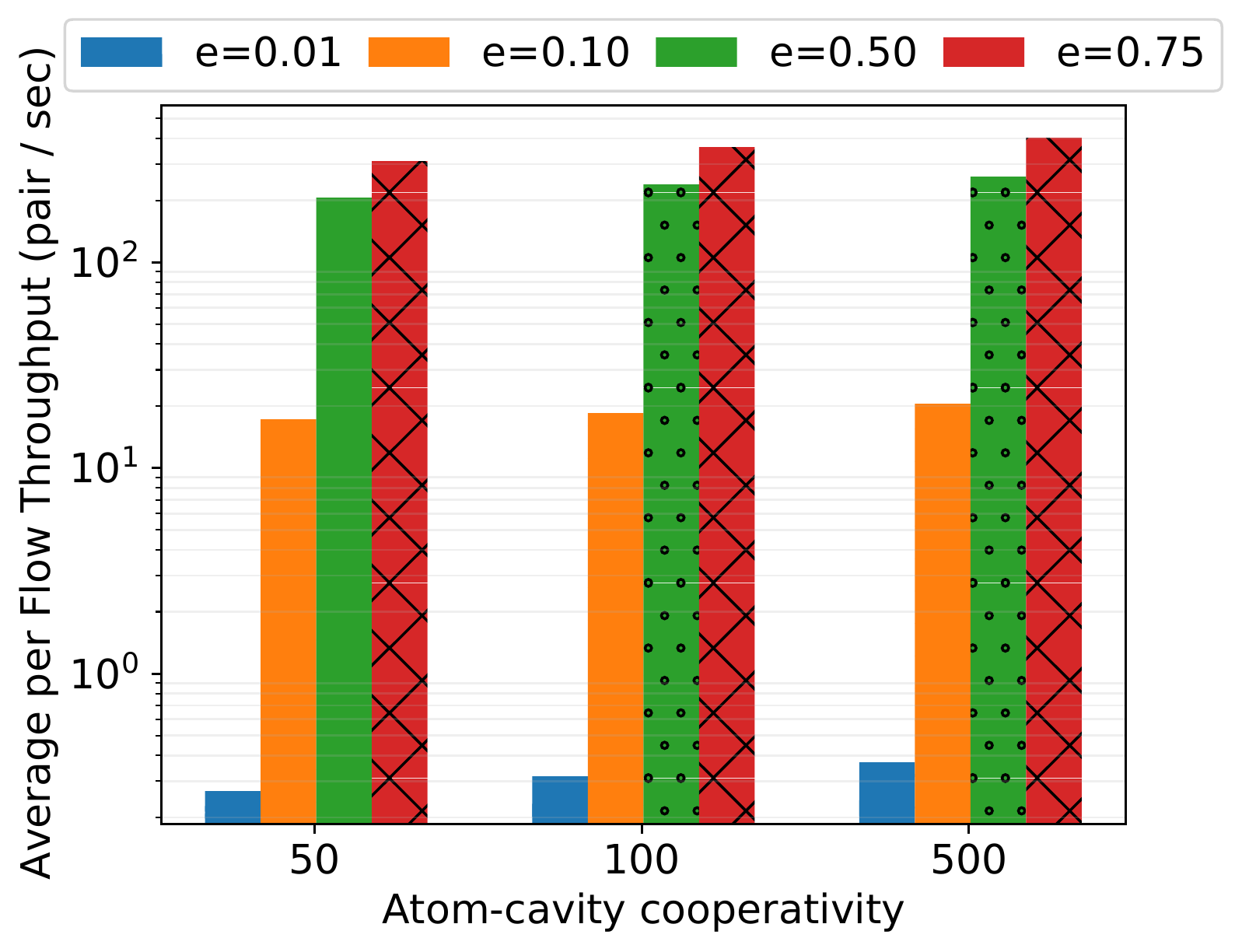}
    \caption{Average per flow throughput for various combinations of memory efficiency and atom-cavity cooperativity.}
    \vspace{-15pt}
    \label{fig:result1}
\end{figure}

Experimental cooperativity $C$ for Er emitters coupled to an optical cavity is in the range of 50 to 500~\cite{Dibos_2018, raha_2020}. Memory efficiencies are not yet perfect and can vary significantly depending on the photonic coupling scheme adopted~\cite{Hensen2015, raha_2020,Zhong_2018}. Therefore, we simulated quantum memories with cooperativity $C = \{50,\allowbreak 100,\allowbreak 500\}$ and efficiency $e = \{0.01,\allowbreak 0.10,\allowbreak 0.50,\allowbreak 0.75\}$. Figure~\ref{fig:result1} shows the average throughput per flow for each of the simulated values of cooperativity $C$ and efficiency $e$. We observe the following. First, performance generally improves with increasing efficiency. This is because a memory with the same cooperativity will generate entanglement with higher probability, reducing the number of entanglement generation protocol failures. Second, the improvement of cooperativity does not provide as much increase in throughput as the improvement of efficiency because for the range of simulated cooperativity the fidelity is already high (>0.9).

This use case illustrates how to use SeQUeNCe to explore the effect of hardware parameters. This particular example is valuable for experimentalists building network prototypes because it identifies the threshold hardware metrics to target useful network throughput.

{}

\subsection{Impact of Classical Channel Delay}
\label{subsec:impact_from_cc}

Quantum routers rely on classical messages to determine the state of quantum memories, even if manipulating quantum states does not inherently require classical messages. The delay of classical channels thus affects the latency of quantum network protocols. Next, we compare the quantum network performance in two scenarios: (1)~using the measured RTT delays in the Chicago topology and (2)~using projected delays of dedicated classical channels that try to avoid unnecessary queuing and processing delays. These projected delays were calculated by adding propagation delay (speed of light in the fiber for the given distance), a transmission delay of 8 \si{\micro\second}, and a processing delay of 4 \si{\micro\second}. The transmission delay corresponds to sending a 1,000-byte control packet over a 1 Gbps link, and the processing delay represents the time to process the packet~\cite{processing_time}. We assume each packet has the highest priority, and we ignore queuing delays. Under these assumptions, the delay for a typical classical channel in our setup reduces from milliseconds to microseconds.

\begin{figure}[]
    \centering
    \includegraphics[width=0.75\linewidth]{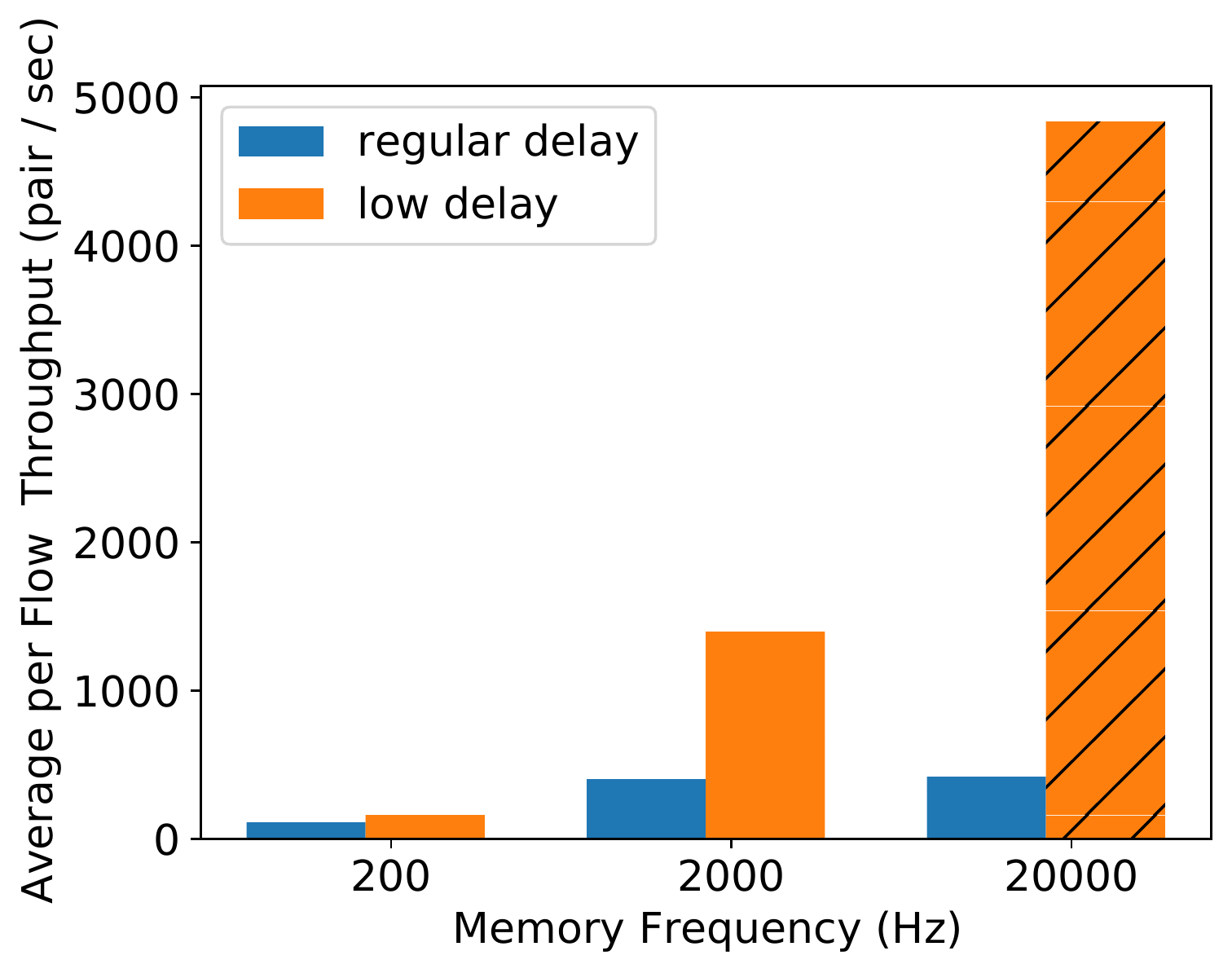}
    \caption{Average per flow network throughput achieved for different quantum memory frequencies.}
    \vspace{-15pt}
    \label{fig:result2}
\end{figure}

To evaluate the effect of this delay reduction, we analyze the performance of the network by recording the network flow throughputs for different memory frequencies. We set the memory efficiency to $e = 0.75$ and the cooperativity to $C = 500$, a combination that achieves the best performance (see \S\ref{subsec:impact-of-qmem}). Our simulation results are shown in Figure~\ref{fig:result2}. We observe that the scenario with low classical channel delay gains more benefit at higher memory frequencies. For example, the average throughput improves by more than a factor of 10 for 20 kHz memories. Another interesting observation is that the increase of frequency from 2 kHz to 20 kHz does not significantly improve the performance of networks with long classical channel delays. The longer delay imposes a limit on the speed at which states can be manipulated by control protocols, rendering the throughput gain from higher memory frequencies insignificant. Thus, we conclude that quantum networks should use low-latency classical control messages to utilize quantum hardware efficiently.

{}

\subsection{Two Memory Distribution Policies}

\begin{figure}[]
    \centering
    \begin{subfigure}[t]{0.45\linewidth}
        \centering
        \includegraphics[width=\linewidth]{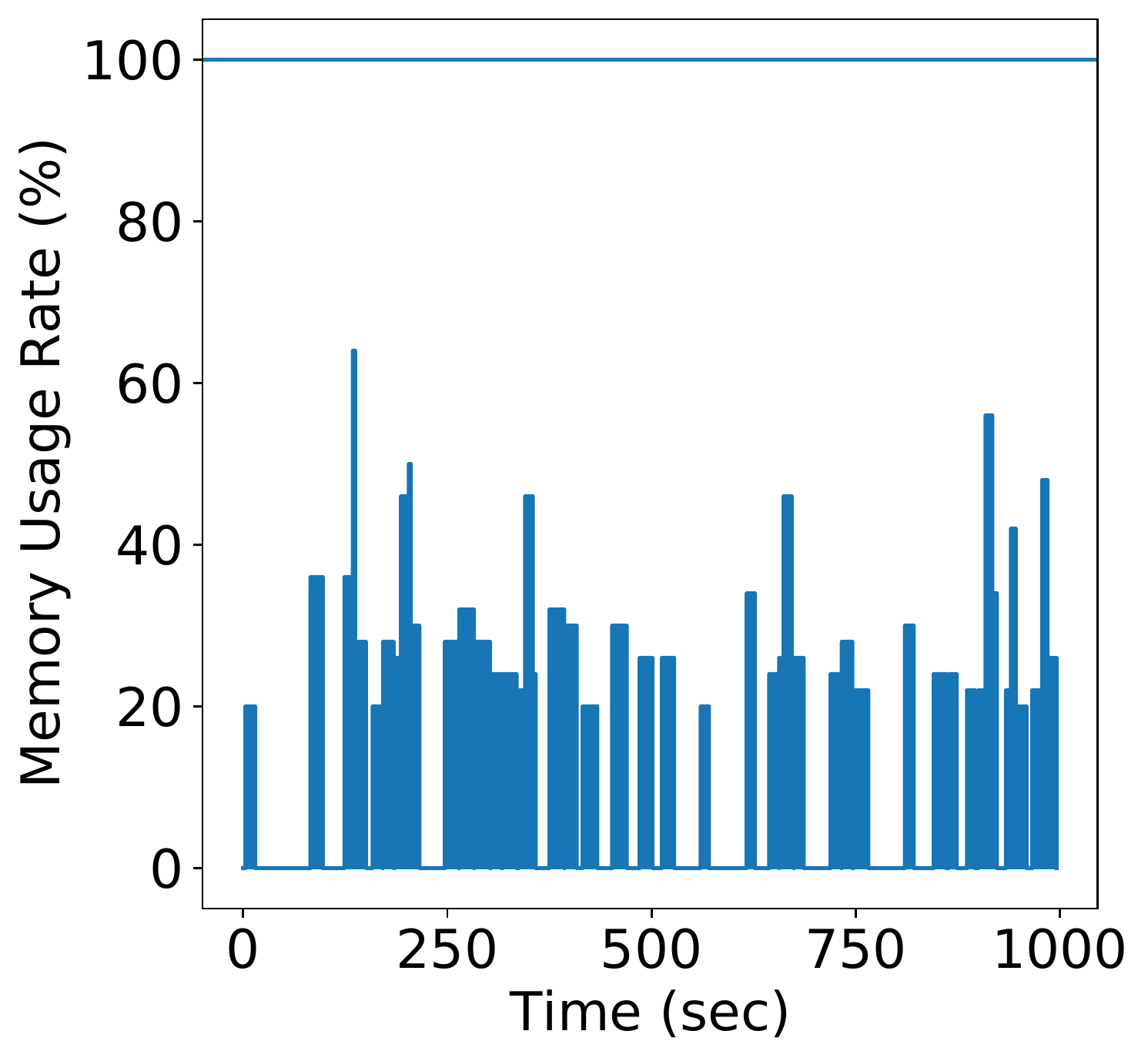}
        \caption{Argonne-3 even}
        \label{fig:before_anl3}
    \end{subfigure}
    \hspace{1em}
    \begin{subfigure}[t]{0.45\linewidth}
        \centering
        \includegraphics[width=\linewidth]{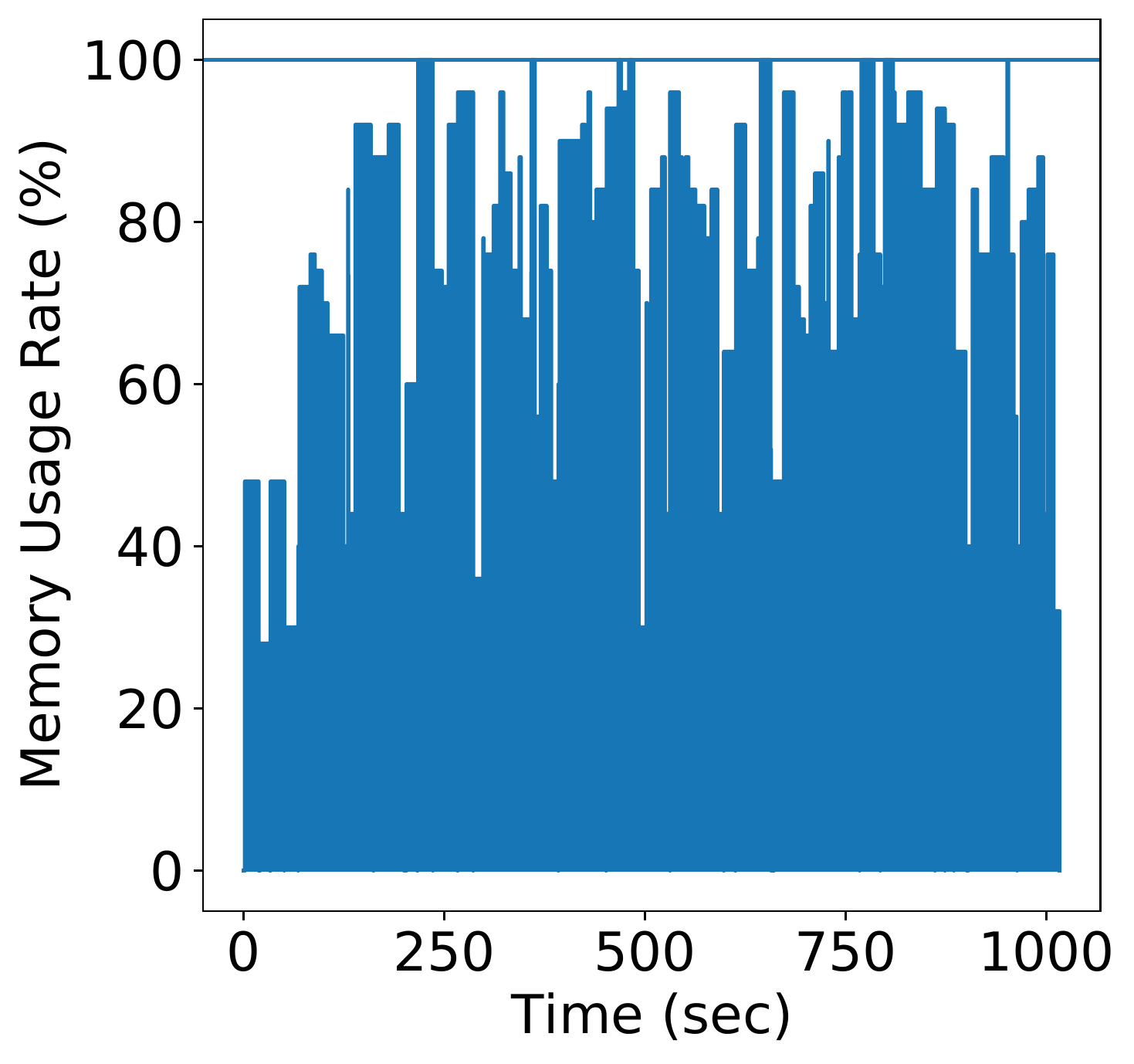}
        \caption{StarLight even}
        \label{fig:before_sl}
    \end{subfigure} 
    \newline
    \begin{subfigure}[t]{0.45\linewidth}
        \centering
        \includegraphics[width=\linewidth]{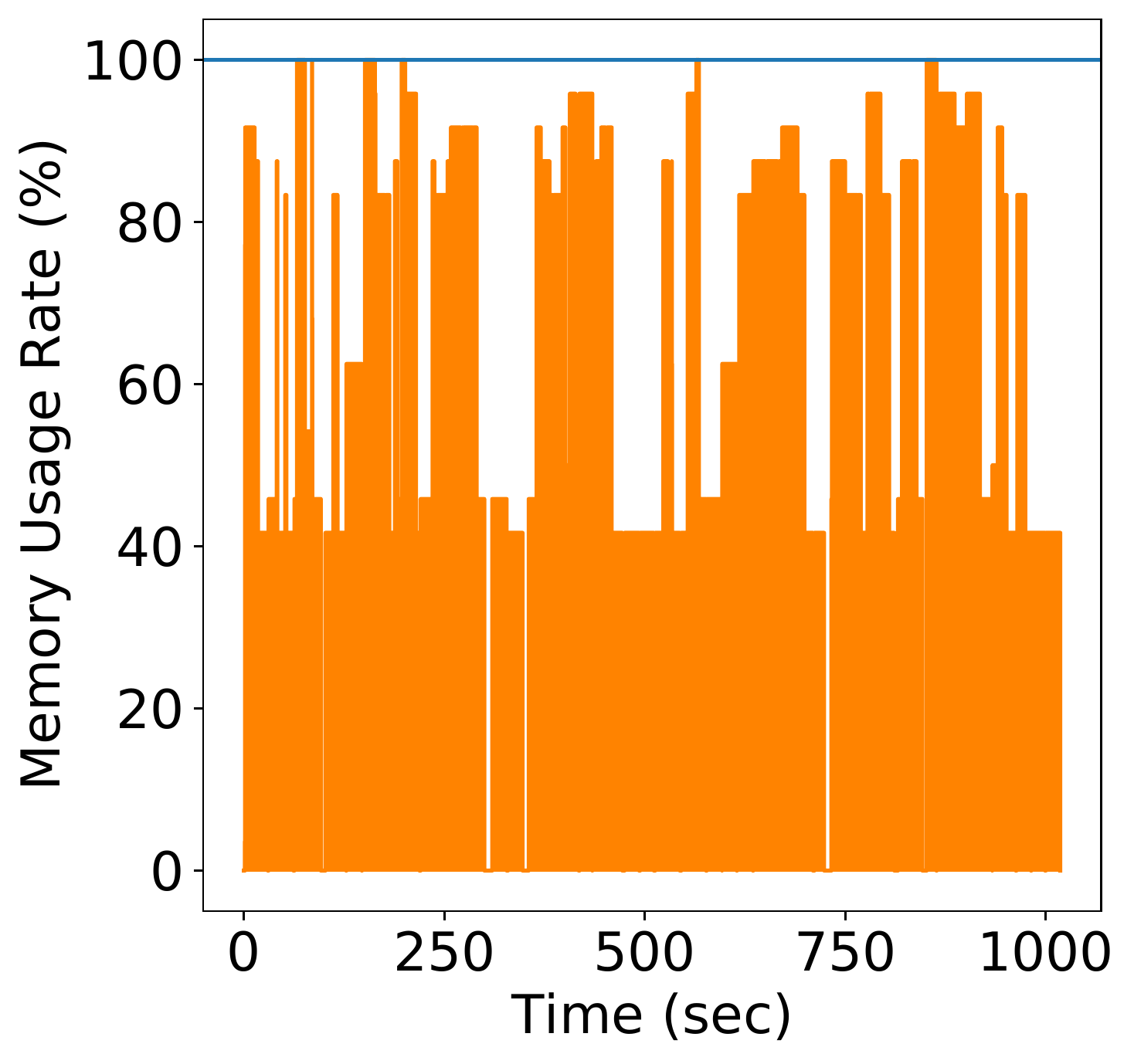}
        \caption{Argonne-3 weighted}
        \label{fig:after_anl3}
    \end{subfigure}
    \hspace{1em}
    \begin{subfigure}[t]{0.45\linewidth}
        \centering
        \includegraphics[width=\linewidth]{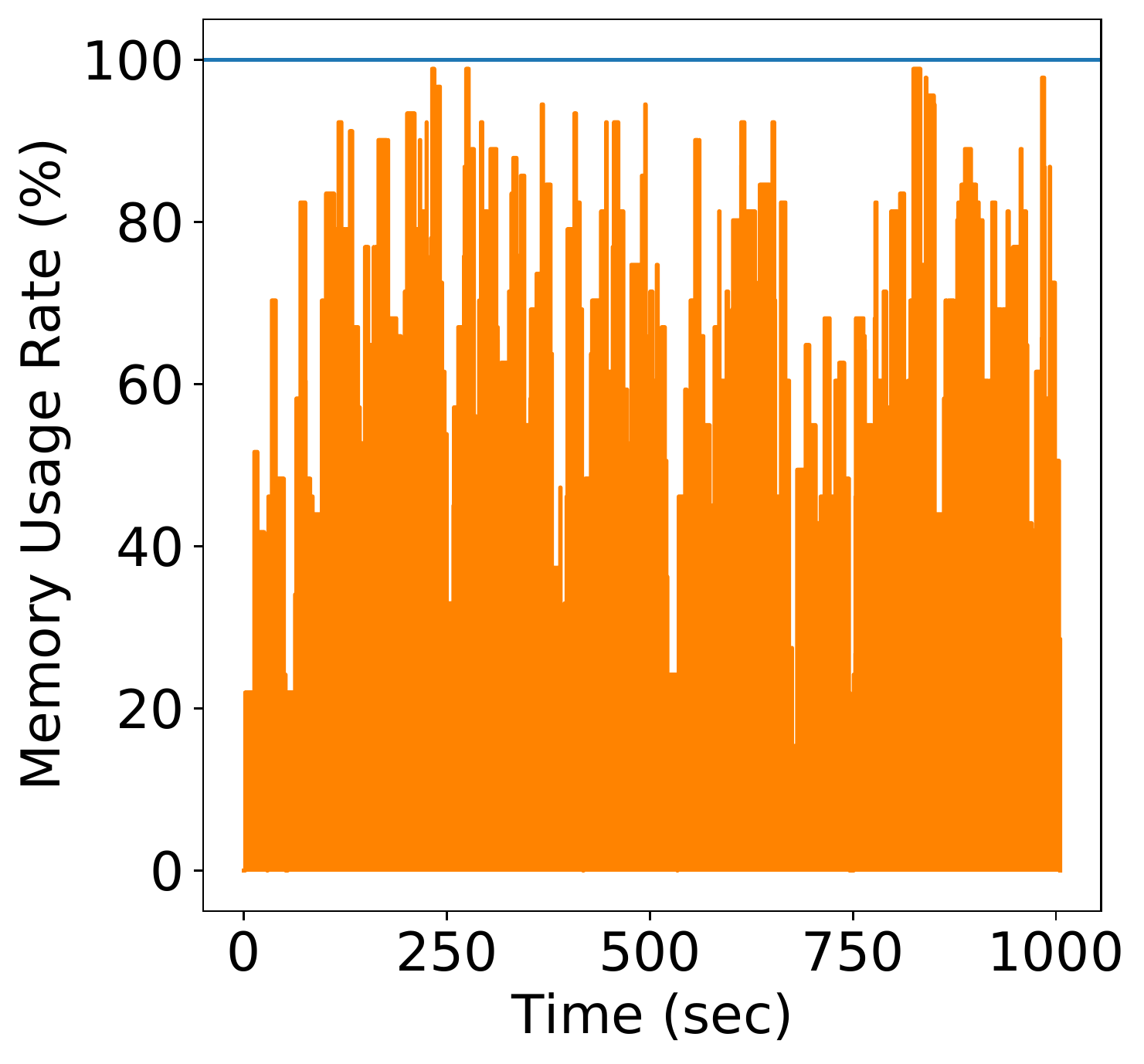}
        \caption{StarLight weighted}
        \label{fig:after_sl}
    \end{subfigure}
    \vspace{-5pt}
    \caption{Memory utilization 
    at Argonne-3 and StarLight nodes. Utilization is shown for even distribution of memory capacities among all nodes (top) and for weighted distribution based on anticipated load (bottom).}
    \vspace{-10pt}
    \label{fig:result3}
\end{figure}

The two previous simulations allocate memory arrays of the same size to each node. However, quantum routers have different workloads depending on their location, connectivity, and application requests. For example, in the Chicago network the StarLight node will experience a heavier workload because it is a router with degree 4 connectivity, whereas Argonne-3 will experience a lower workload as an end host. The memory utilization history for the Argonne-3 and StarLight nodes in a scenario where each of the 9 network nodes uses a memory array of size 50 is shown in Figures~\ref{fig:before_anl3} and~\ref{fig:before_sl}, respectively. Not surprisingly, the memory usage rate in the Argonne-3 node is low whereas the memory usage rate in the StarLight node often approaches 100\%.

Clearly, given a fixed budget of memory arrays, the overall network throughput can be improved by redistributing some of the memories between the nodes. Here we repeat the previous simulation after redistributing the memories according to the anticipated load. This was achieved by considering all possible routes in the network and counting the incidence of network nodes over all routes. Our RSVP protocol requires double the memory in intermediate nodes compared with the Initiator and Responder nodes. Therefore, when we consider node incidence on a route, intermediate nodes carry twice the weight. The weighted assignment of memory arrays using this algorithm for a fixed memory budget of size 450 (the same memory budget that was used for the even distribution policy) on the Chicago network is shown in Table~\ref{tab:memo_dist}.

\begin{table}[]
\caption{Memory array size distributed by usage.}
\vspace{-5pt}
\begin{tabular}{|l|l|l|l|}
\hline
\textbf{Argonne-1}     & 103   & \textbf{NU-Evanston}  & 25 \\ \hline 
\textbf{Argonne-2}     & 25    & \textbf{StarLight}    & 91 \\ \hline
\textbf{Argonne-3}     & 24    &\textbf{UChicago-HC}   & 24  \\ \hline
\textbf{Fermilab-1}    & 67    & \textbf{UChicago-PME} & 67 \\ \hline
\textbf{Fermilab-2}    & 24    &                       &    \\ \hline
\end{tabular}
\label{tab:memo_dist}
\vspace{-10pt}
\end{table}

Figures~\ref{fig:after_anl3} and~\ref{fig:after_sl} show the usage rate of memories in the Argonne-3 and StarLight nodes with the new weighted distribution of quantum memories. Compared with distributing memories evenly, distributing memories by weight increases the average usage rate at Argonne-3 from 17.4\% to 56.0\% and at StarLight from 53.3\% to 56.6\%, leading to a more even use of this limited resource while allowing the service of more requests. The percentage of time when the memory arrays in the StarLight node are more than 90\% utilized decreased from 43.5\% to 24.4\%, reducing the occurrence of rejected application requests due to insufficient memory at the highly connected StartLight router node.

\begin{figure}[]
    \centering
    \begin{subfigure}[t]{0.45\linewidth}
        \centering
        \includegraphics[width=\linewidth]{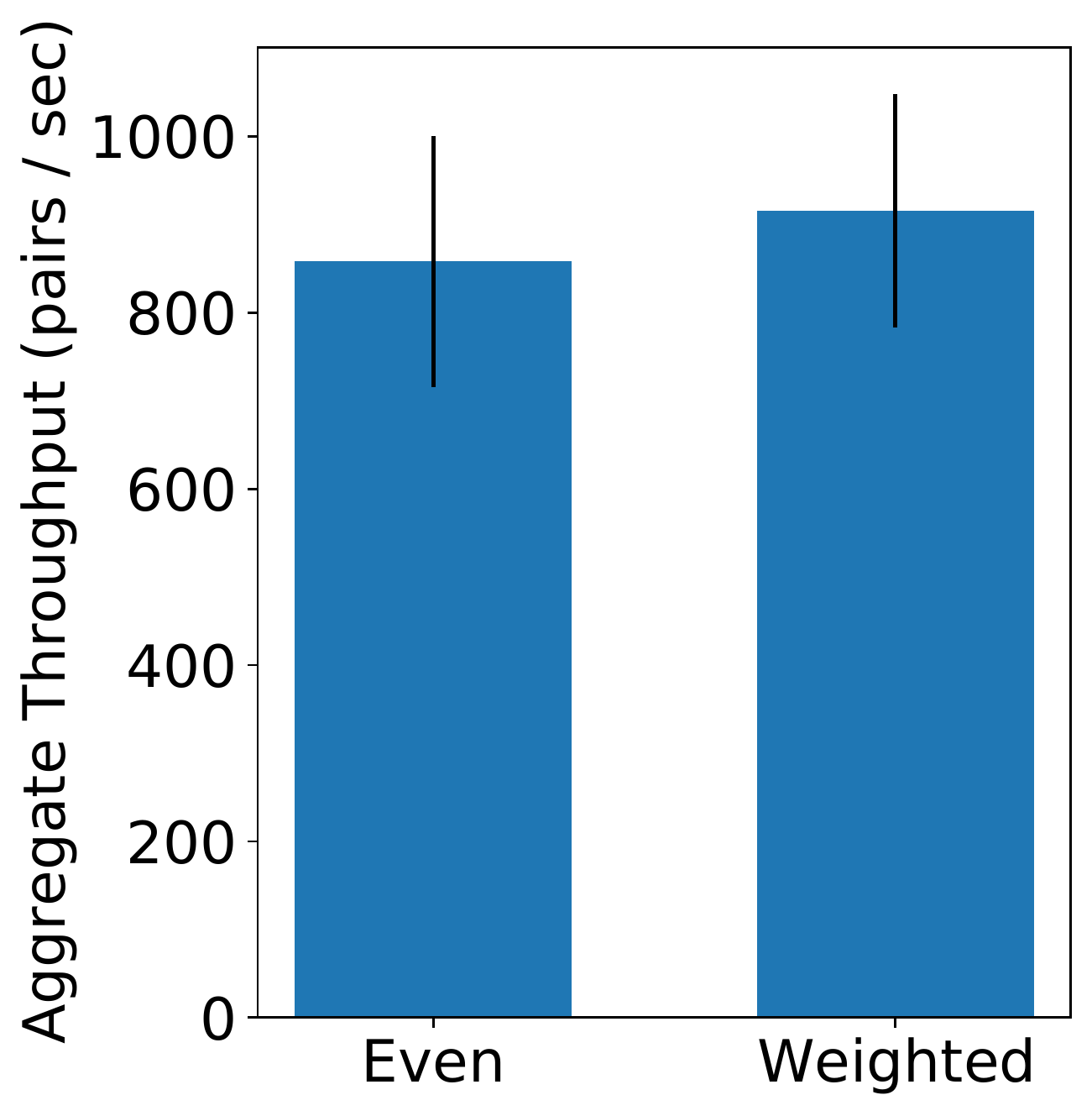}
        \caption{Average network throughput}
        \label{fig:dist_ent}
    \end{subfigure}
    \hspace{1em}
    \begin{subfigure}[t]{0.45\linewidth}
        \centering
        \includegraphics[width=\linewidth]{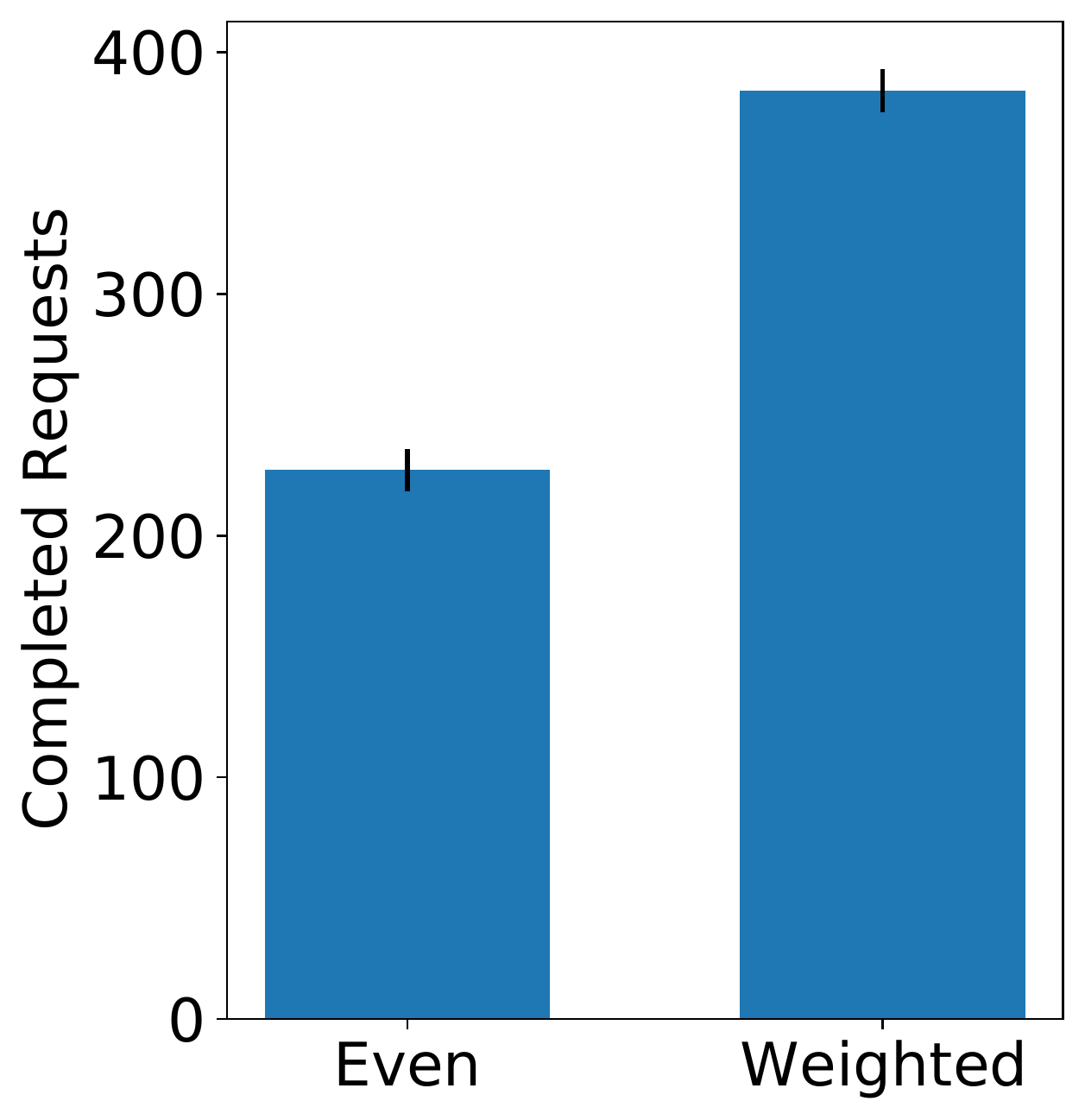}
        \caption{Total number of completed requests}
        \label{fig:app_req}
    \end{subfigure}
    \vspace{-5pt}
    \caption{Average aggregate network throughput and number of completed requests for the two memory distribution policies.}
    \vspace{-15pt}
    \label{fig:result4}
\end{figure}

We repeated simulations for both memory allocation policies ten times with different random seeds and recorded the aggregate network throughput and completed number of requests. The results are shown in Figure~\ref{fig:result4}. We observe that the weighted memory distribution policy improved the aggregate network throughput by 6.8\% and the number of completed requests by 69.1\% on average. Given a fixed budget of memories, the capacity of a quantum network can thus be improved by adjusting the memory distribution policy. 

\section{Related Work}
\label{sec:relatedwork}

Besides SeQUeNCe, there are two other quantum network simulators with ambition to comprehensively model the behavior of full-stack quantum networks. QuISP \cite{matsuo2019quantum} has been introduced as an open source package in mid-2020, and NetSquid~\cite{link-layer-protocol} recently became available upon request. QuISP~\cite{matsuo2019quantum} supports various entanglement purification protocols, link tomography, and entanglement swapping. Published simulations~\cite{matsuo2019quantum} focus on faithful error modeling but restrict the studied scenarios to point-to-point communication with and without repeater nodes over distances of up to 50 km. Implementation of higher layers including routing and more complex network topologies are planned for future work. According to the NetSquid~\cite{NetSquidWebsite} project website, it generates and processes quantum information using nitrogen-vacancy centers in diamond. In comparison, SeQUeNCe uses single rear-earth ion memories, functionality that not available in NetSquid. Work published by the NetSquid team~\cite{matsuo2019quantum} focuses on development and analysis of physical and link layer protocols, and evaluations focus on simulations of point-to-point communication between two nodes in the lab (2 m) and two cities (25 km).

All three simulators are quickly evolving and features are added on a daily basis, making comparisons quickly obsolete. The biggest differences can be found in architectural assumptions. For example, QuISP uses RuleSets that are distributed to nodes along a path at connection setup time. NetSquid proposes to use a 5-layer quantum network stack with Physical, Link, Network, Transmission, and Application layers. SeQUeNCe follows a modularized design with cross-module communication to allow maximum flexibility.

Recently introduced, QuNetSim~\cite{qunetsim} is a simulator capable of smaller-scale simulations of 5 to 10 nodes that also uses a network layering framework inspired by the OSI model. However, the simulator  focuses only on the upper layers, leaving the physical layer unspecified. QuNetSim also does not allow simulating repeater nodes and the associated protocols that are required for long-distance communication. Another simulator, SQUANCH~\cite{SQUANCH}, is notable for its agent-based modeling, allowing natural parallelization by running each agent on its own process. Because the simulator does not attempt to model interactions of protocols or details at the physical layer, the simulator is suitable for demonstrations of concepts such as teleportation~\cite{Teleportation} or superdense coding~\cite{SuperdenseCoding}. SimulaQron~\cite{simulaqron} does not aim to simulate quantum networks but it facilitates development network applications. It uses the classical-quantum combiner interface~\cite{cqc} with sockets that mimic information transmission over simulated quantum channels.

Many additional simulation tools exist that can be used to perform numeric studies of individual quantum network algorithms and protocols~\cite{qkdsim, Hu2019, ns3_qkd, qcp_simulate, pereszlenyi2005simulation}. Many additional tools are listed at the website of the Quantum Open Source Foundation~\cite{QOSF}.

\section{conclusion}
\label{sec:conclusion}

In this paper we introduced SeQUeNCe, a customizable discrete-event quantum network simulator. We introduced a modularized design for the simulator that closely matches an abstracted quantum network architecture, and we implemented a high-performance simulation kernel that allows simulating transmission and tracking of photon pulses and control messages with picosecond accuracy. We also implemented a comprehensive suite of quantum network protocols, including translation of the algorithmic descriptions into fully functional protocol state machines.

Our planned next steps include support for additional protocols to allow more comprehensive performance evaluations, as well as parallelization of the tool for use on multinode, multicore supercomputers. By releasing SeQUeNCe as open source software, we aim to generate community interest in using and extending the simulator, as well as allow comparisons of experimental and simulation results obtained by other teams.

\begin{acks}
We thank Eugene Wang and Siu Man Chan for contributions to the SeQUeNCe software package. We also thank Liang Jiang and Manish Kumar Singh for valuable feedback. This material is based upon work supported by Laboratory Directed Research and Development (LDRD) funding from Argonne National Laboratory, provided by the Director, Office of Science, of the U.S. Department of Energy under contract DE-AC02-06CH11357. 
\end{acks}

\bibliographystyle{ACM-Reference-Format}
\bibliography{acmart.bib}

\section*{Appendices}
\appendix

\section{Barrett-Kok Protocol} \label{apdx:barret}

The Barrett-Kok generation protocol~\cite{barrett_eg} is defined as follows:

\begin{enumerate}[leftmargin=*]
    \item Excite each quantum memory. The quantum state of the system transforms according to $| + \rangle \otimes | + \rangle \rightarrow \frac{| \uparrow \rangle + | e \rangle}{\sqrt{2}} \otimes \frac{| \uparrow \rangle + | e \rangle}{\sqrt{2}}$
    \item Expect a photon detection event in either $D^+$ or $D^-$ during a time window $t_{wait}$. If both detectors are (not) clicked, the quantum state of system is $| ee \rangle$ ($|\uparrow \uparrow \rangle$). The scheme has then failed, and the qubits must be newly prepared before reattempting the entangling procedure.
    If only one detector is triggered, the quantum state still cannot be determined as a maximal state; 
    state $| ee \rangle$ can also produce the same phenomenon if one photon is lost. Therefore, the current quantum state is $\frac{| \uparrow e \rangle + | e \uparrow \rangle + | ee \rangle}{\sqrt{3}}$
    \item Memories wait a further time $t_{relax}$ for the transformation of memory states $| e \rangle \rightarrow | \downarrow \rangle$. The state of the system is $\frac{| \uparrow \downarrow \rangle + | \downarrow \uparrow \rangle + | \downarrow \downarrow \rangle}{\sqrt{3}}$
    \item Apply an X-gate (a quantum gate) to both memories. After the X-gate, the quantum state of the memories is $\frac{| \downarrow \uparrow \rangle + | \uparrow \downarrow \rangle + | \uparrow \uparrow \rangle}{\sqrt{3}}$.
    \item Excite each quantum memory again. The state of the system is $\frac{| e \uparrow \rangle + | \uparrow e \rangle + | \uparrow \uparrow \rangle}{\sqrt{3}}$
    \item Expect a photon detection event in either $D^+$ or $D^-$ within a time $t_{wait}$ window. If only one detector is triggered, the maximally entangled state $\frac{|\downarrow \uparrow \rangle + |\uparrow \downarrow \rangle}{\sqrt{2}}$ is established. Otherwise, the scheme has failed. If the detection from each round is observed in the same (different) detector, the final state is $| \Psi^+ \rangle$ ($| \Psi^- \rangle$). Then the protocol ends.
\end{enumerate}

\begin{figure}[b]
    \centering
    \includegraphics[width=0.85\linewidth]{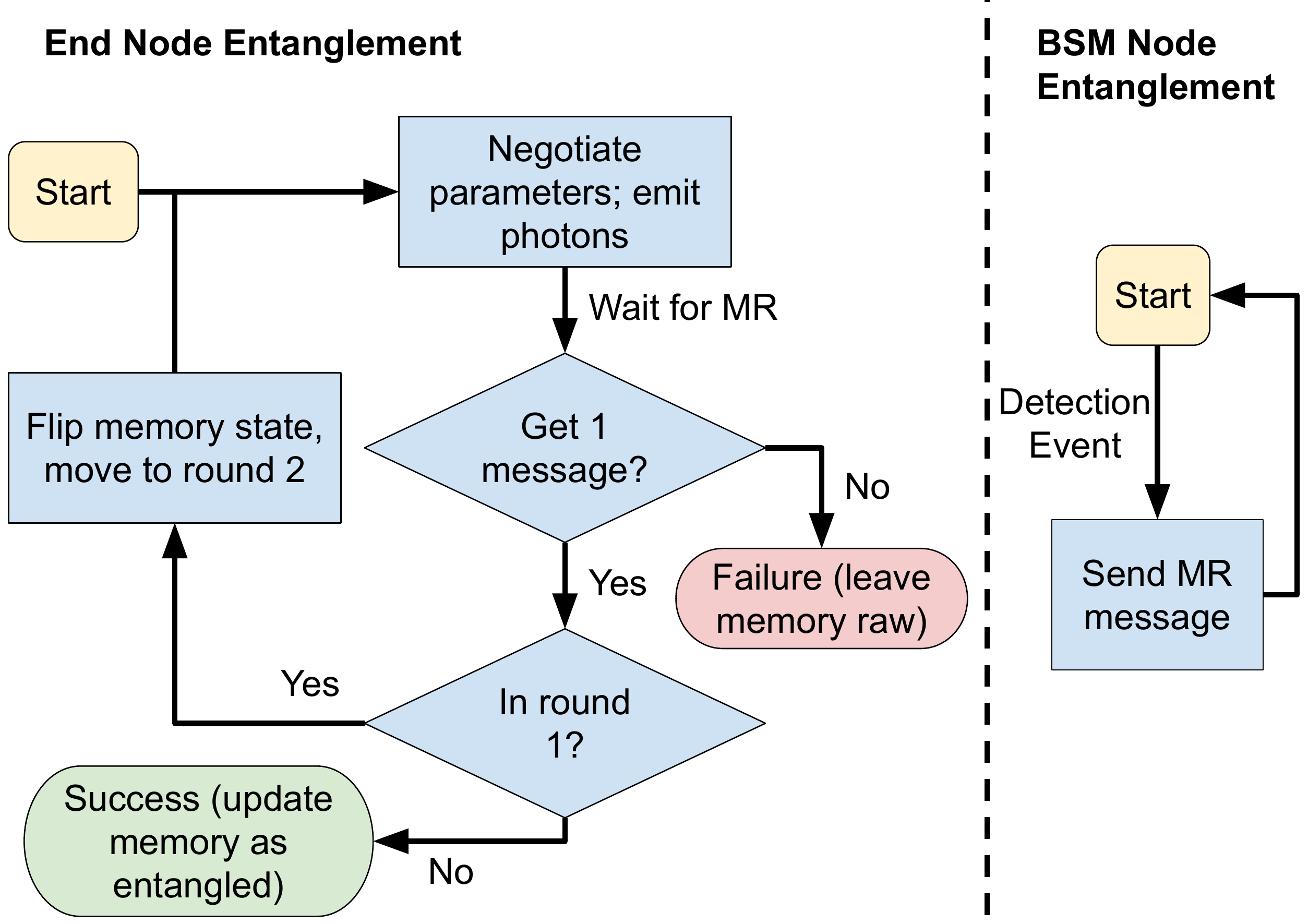}
    \caption{State machine of entanglement generation.}
    \label{fig:state_machine1}
\end{figure}{}

The SeQUeNCe implementation of the Barrett-Kok protocol is instantiated on each quantum node and paired by the Resource Management module. The state machines of the protocol instances at the quantum nodes and at the BSM are shown in Figure~\ref{fig:state_machine1}. We now describe the classical message exchange between a pair of nodes using the Barrett-Kok protocol in SeQUeNCe. A designated member of the pair will start communication with a \texttt{NEGOTIATE} message (shown as ``negotiate parameters'' in Figure~\ref{fig:state_machine1}). The protocol instance on the other node responds with an offer, and the Barrett-Kok protocol can begin. The \texttt{NEGOTIATE} message and its response ensure that the two memories emit photons simultaneously. The two quantum nodes then schedule the \texttt{excite} operations on their respective memories and wait for a measurement result via the \texttt{MR} (measurement result) message from the the BSM node. If an improper result is received, the protocol ends with a failure, and each instance reports its failure to its respective Resource Manager. Otherwise, the protocol instance applies an X-gate and proceeds to round 2, repeating the process of round 1. If both rounds are completed successfully, the two memories are confirmed to be entangled, an entanglement result is returned to the Resource Manager. Regardless of whether the protocol succeeds or fails, the protocol instance destroys itself and releases the quantum memories back to the Resource Manager. The protocol instance at the node executing the BSM is not created or destroyed by a top-level resource manager; it is always present. It waits in an idle start state for any photon detection event and then notifies adjacent quantum nodes of the entanglement result with a \texttt{MR} message. Note that the lifetime of entanglement starts at the first \texttt{excite} operation.

\section{BBPSSW Purification Protocol} \label{apdx:purification}

The BBPSSW~\cite{bbpssw} protocol is a purification protocol that uses two pairs of qubits $A_1B_1$ and $A_2B_2$ and assumes these pairs of qubits have the same fidelity $F > 0.5$. The protocol consists of the following steps: (1) apply a local CNOT operations $U_{CNOT}^{A_1 \rightarrow A_2} \otimes U_{CNOT}^{B_2 \rightarrow B_1}$; (2) measure qubit $A_2$ ($B_2$) in Z-basis (X-basis) with corresponding results $(-1)^{\zeta_1}$ ($(-1)^{\xi_1}$), respectively, where $\zeta_1, \xi_1 \in \{0,1\}$; and (3) discard the measured pair $A_2B_2$ and keep the purified pair $A_1B_1$ if $(\zeta_1 + \xi_1) mod 2 = 0$.

\begin{figure}[htbp]
    \centering
    \includegraphics[width=0.68\linewidth]{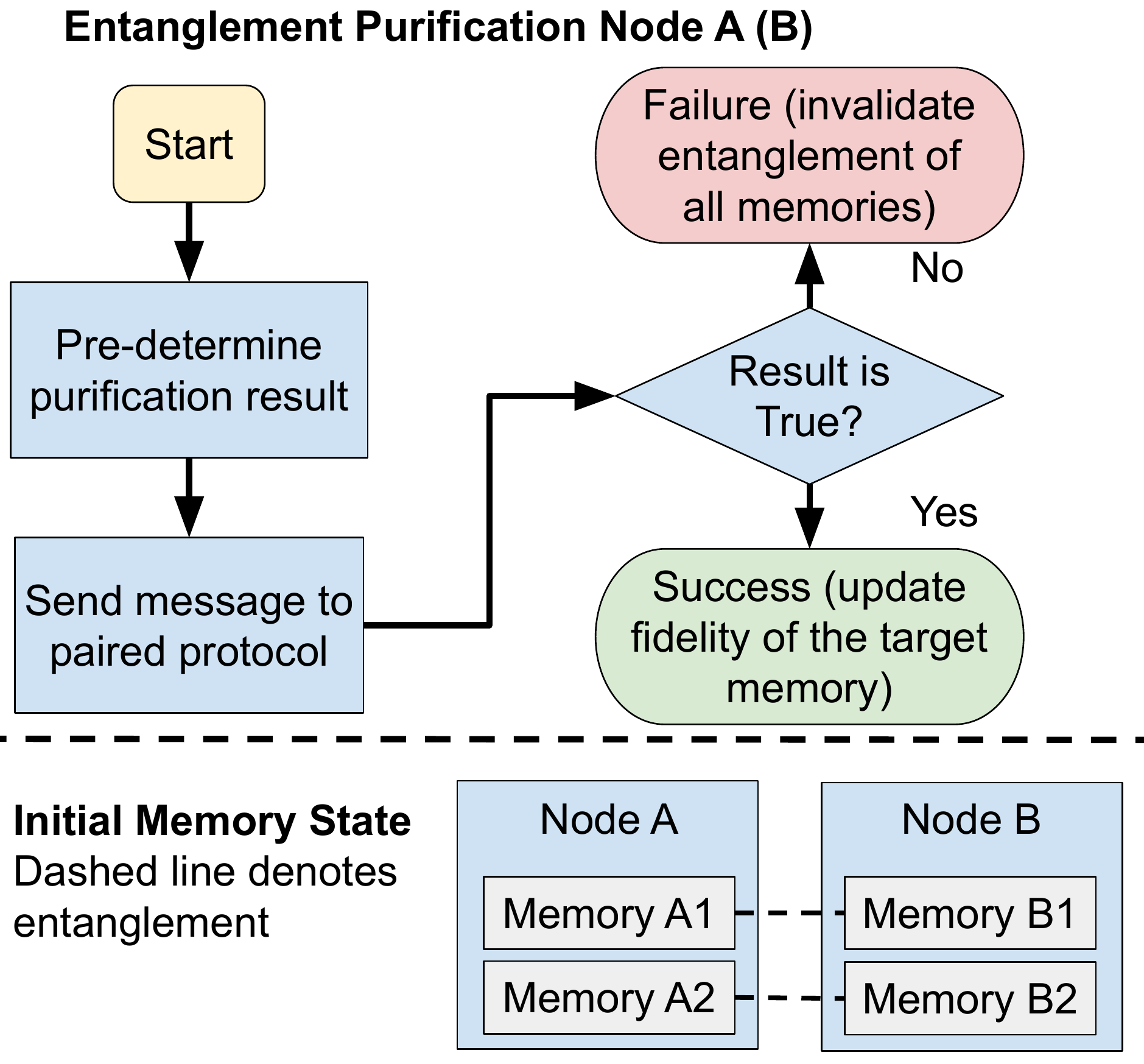}
    \caption{State machine of entanglement purification.}
    \label{fig:state_machine2}
\end{figure}

The instances of the modeled BBPSSW protocol are created and paired by their respective Resource Managers. When two of these instances are initialized properly, they start the procedure of purification. Figure~\ref{fig:state_machine2} shows the state diagram for the BBPSSW protocol model. Based on the fidelity of the two given memories, the protocol instances determine the probability of success $p_{suc}$ and randomly determine the success of purification with this probability (but do not update the state of the memories). Then, the two BBPSSW instances send messages to each other to mimic the transmission of measurement results and wait for the message from the other side. When a protocol instance receives a message, the predetermined result of purification is used to update the state of the quantum memory, and the measured memory loses its entangled state. If the predetermined result indicates the success of purification, the target memory keeps its entanglement state and improves fidelity. Otherwise, the target memory also loses its entanglement.

{}

\section{Entanglement Swapping} \label{apdx:es}

\begin{figure}[htbp]
    \centering
    \includegraphics[width=0.85\linewidth]{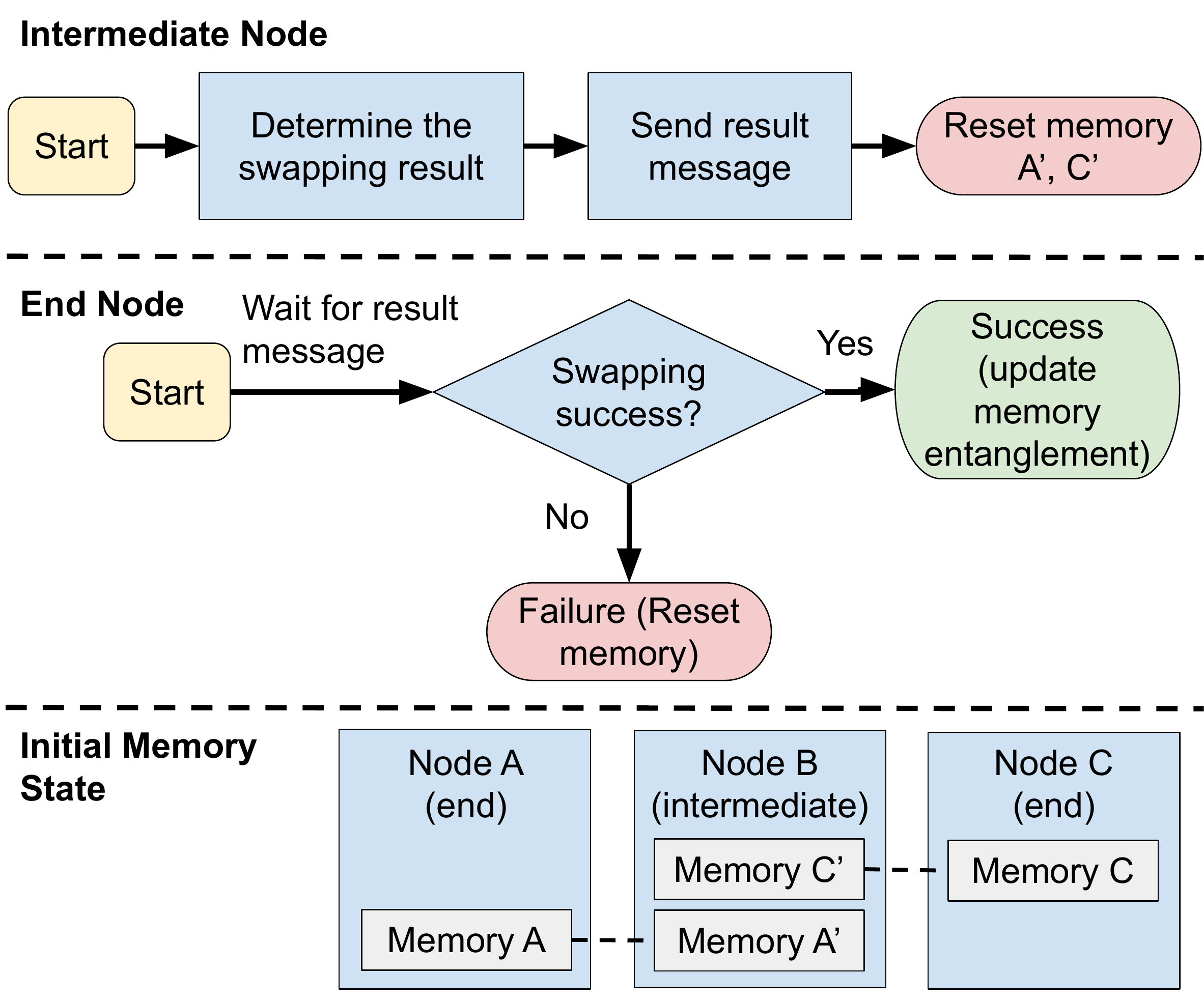}
    \caption{State machine of entanglement swapping.}
    \label{fig:state_machine3}
\end{figure}

Figure~\ref{fig:state_machine3} shows the state machines for the entanglement swapping protocol at both end and intermediate nodes. The intermediate node uses the probability of success and random number to determine whether swapping succeeds and sends a message to the paired end nodes. Then, the intermediate finishes its work and releases resources. The end nodes start and wait for the message from the intermediate. If the message shows the swapping succeeded, the protocol instance updates the identity of the entangled memory and fidelity. Otherwise, the entanglement state is removed. The updated memory is then released to the Resource Manager.

{}

\section{Reservation Protocol} \label{apdx:reservation}

\begin{figure}[htbp]
    \centering
    \includegraphics[width=\linewidth]{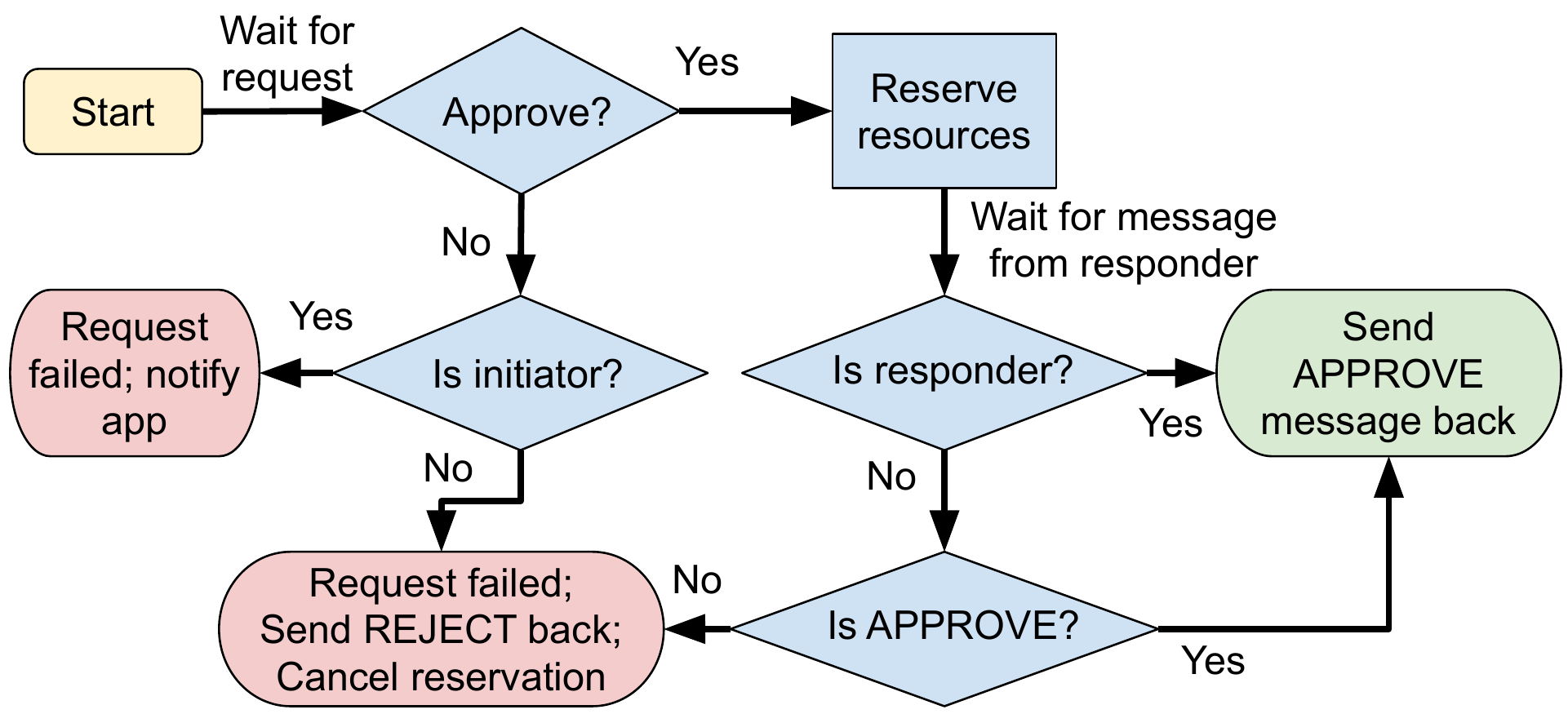}
    \caption{State machine of reservation protocol.}
    \label{fig:reservation_state_machine}
\end{figure}

Figure~\ref{fig:reservation_state_machine} shows the state machine of the reservation protocol. A protocol instance starts upon arrival of a request message.
The protocol instance checks whether sufficient local memory is available from the start time to the end time of the request. If sufficient memory is available, the instance will attach its local information and forward the message to the next hop in the path until the message arrives at the Responder.
Every quantum router in the path holds the quantum memory resources until it receives an \texttt{APPROVE} message.
When the request is approved by the Responder, this \texttt{APPROVE} message is created and sent back to confirm the success of reservation. Once a protocol instance confirms reservation with an \texttt{APPROVE} message, the corresponding local rules are created and scheduled. If the protocol instance of any node on the path (Responder included) rejects the request, a \texttt{REJECT} message is sent back to the previous hop until it arrives at the Initiator. All reservations from the request are canceled. When the protocol instance on the Initiator receives the result of the request, it sends the result to the Application module. 

{}

A reservation protocol instance creates three rules to provide service. The first rule defines the conditions and actions for entanglement generation. If the state of reserved memory is \texttt{raw}, the memory is allocated to an instance of the generation protocol. The second rule defines conditions and actions for entanglement purification. If two reserved memories have same fidelity of entanglement and the current fidelity is lower than the target fidelity, the two memories are allocated to an instance of the purification protocol. The third rule defines conditions and actions for entanglement swapping. If the two reserved memories are entangled with memories on specific nodes and the fidelity of entanglement is larger than the target fidelity, the two memories are allocated to an instance of the entanglement swapping protocol. 

\end{document}